\documentclass[structabstract]{aa}  
\usepackage{natbib}
\usepackage[outdir=./]{epstopdf}
\bibpunct{(}{)}{;}{a}{}{,}
\usepackage{lscape}
\usepackage{rotating}
\usepackage{color}
\usepackage[table]{xcolor}

\usepackage{lipsum}
\usepackage{graphicx}
\usepackage{enumerate}
\usepackage{siunitx}
\usepackage{multirow}
\usepackage{tabularx}
\usepackage{txfonts}
 \def\mso{\,\mathrm{M}_\odot}
 \def\rso{\,{\rm R}_\odot}

 \def\egs{\, {\rm erg}\, {\rm s}^{-1}}

 \def\simle{\mathrel{\hbox{\rlap{\hbox{\lower4pt\hbox{$\sim$}}}\hbox{$<$}}}}
 \def\simgr{\mathrel{\hbox{\rlap{\hbox{\lower4pt\hbox{$\sim$}}}\hbox{$>$}}}}
 \def\msoy{\, \mso~{\rm yr}^{-1}}

\begin{document}
  \title{Nuclear timescale mass transfer in models of supergiant and ultra-luminous X-ray binaries}

   \author{M. Quast \inst{1}
          \thanks{e-mail: mquast@astro.uni-bonn.de},
          N. Langer \inst{1,2}
          \and
          T. M. Tauris \inst{3,4}
          }

   \authorrunning{Quast et al.}
   \titlerunning{Nuclear timescale mass transfer in supergiant X-ray binaries}

   \institute{Argelander-Insitut f\"ur Astronomie, Universit\"at Bonn, Auf
              dem H\"ugel 71, 53121 Bonn, Germany
              \and
              Max-Planck-Institut f\"ur Radioastronomie, Auf dem Hügel 69, 53121 Bonn, Germany
              \and
              Aarhus Institute of Advanced Studies (AIAS), Aarhus University, H{\o}egh-Guldbergs~Gade~6B, 8000~Aarhus~C, Denmark
              \and
              Department of Physics and Astronomy, Aarhus University, Ny~Munkegade~120, 8000~Aarhus~C, Denmark
}

\date{Draft version March 12, 2019}

\abstract{The origin and number of the Galactic supergiant X-ray binaries is currently not well understood.
They consist of an evolved massive star and a neutron star or black hole companion. X-rays are thought to be generated 
from the accretion of wind material donated by the supergiant, while mass-transfer due to Roche-lobe overflow is mostly 
disregarded, since the high mass ratios of these systems is thought to render this process unstable. 
}  
{We investigate how the proximity of supergiant donor stars to the Eddington-limit, and their advanced evolutionary 
stage, may influence the evolution of massive and ultra-luminous X-ray binaries with supergiant donor stars (SGXBs and ULXs).} 
{We construct models of massive stars with different internal hydrogen/helium gradients and different hydrogen-rich envelope masses, 
and expose them to slow mass loss to probe the response of the stellar radius. 
In addition, we compute the corresponding Roche-lobe overflow mass-transfer evolution 
with our detailed binary stellar evolution code, approximating the compact objects as point masses.
}
{We find that a hydrogen/helium gradient in the layers beneath the surface, as it is likely present 
in the well-studied donor stars of observed SGBXs, can enable nuclear timescale mass-transfer in SGXBs with a BH or a NS
accretor, even for mass ratios in excess of 20. In our binary evolution models, the donor stars
rapidly decrease their thermal equilibrium radius and can therefore cope with the inevitably strong orbital 
contraction imposed by the high mass ratio. We find the orbital period derivatives of our models
in good agreement with empirical values.    
We argue that the SGXB phase may be preceded by a common envelope evolution.
The envelope inflation near the Eddington-limit makes this mechanism more likely to occur at high metallicity.
}
{
Our results open a new perspective for understanding the large number of Galactic SGXBs, and their almost complete absence in the SMC.
They may also offer a way to obtain more ULX systems, to find 
nuclear timescale mass-transfer in ULX systems even with neutron star accretors, and shed new light on
the origin of the strong B-field in these neutron stars. 
}
\maketitle
%

\section{Introduction}\label{sec:intro}

X-ray binaries represent an evolved stage of the evolution of massive binary systems
\citep{1992fxra.conf...57V, 2006csxs.book..623T, 2017A&A...604A..55M}. 
They contain an ordinary star, the mass donor, and a compact object, namely a neutron star or a black hole. 
In these systems X-ray radiation is released by the accretion of matter released by the mass donor 
onto the compact companion \citep{1985apa..book.....F}. Depending on the donor star's mass, 
X-ray binaries are divided into low-mass (LMXBs) and high-mass X-ray binaries (HMXBs). 

HMXBs may also help us to understand 
the rare ultra-luminous X-ray sources (ULXs) \citep{2017ARA&A..55..303K}, their more luminous cousins 
in the X-ray sky. 
Furthermore HMXBs may be progenitors of merging back holes \citep{2016A&A...588A..50M} 
and neutron stars \citep{2017ApJ...846..170T}, and thus have a direct connection to 
the gravitational wave signals detected by LIGO and Virgo \citep{2016PhRvL.116f1102A, 2017ApJ...848L..12A}.
A study of HMXBs, their formation, their evolution and their fate, thus provides a 
better understanding of future results from gravitational wave surveys.

While the mass-transfer and accretion processes in LMXBs are quite well understood \citep{2006csxs.book..623T},
our knowledge of that in HMXBs is more limited.
The mass-transfer mode in HMXBs is thought to be either wind accretion \citep{2014EPJWC..6402001S} 
or Roche-lobe overflow (RLO) \citep{1978A&A....62..317S}. Since the first mode requires an 
extreme stellar wind with mass-loss rates of the order of several $10^{-5}\,\text{M}_\sun \text{yr}^{-1}$ 
to achieve the observed X-ray luminosities, one expects this only to happen in HMXBs. RLO is often referred to 
in LMXBs, since it provides a sufficiently large mass-transfer rate to explain the luminous X-ray emission. 
In HMXBs however, RLO is expected to lead to a rapid shrinking of the orbit as the result of the 
large mass ratio between the donor star and the accretor, leading to a common envelope (CE) phase. 
\cite{2017arXiv170102355V} argued that systems with a mass ratio $\gtrsim 3.5$ would always undergo 
unstable RLO mass-transfer. Hence, a system containing an O\,star ($\gtrsim 20\,\text{M}_\sun$) 
and a neutron star ($< 3 \,\text{M}_\sun$) should quickly enter a CE phase. In this case, the resulting X-ray 
lifetime cannot exceed the thermal timescale of the donor star. \cite{1978A&A....62..317S} found indeed an 
X-ray lifetime of only $ 3\times 10^4 \, \text{yr}$ for a system with a mass ratio of 16, 
which is of the order of the O\,star's thermal timescale.

HXMBs are subdivided into two main groups \citep{1995xrbi.nasa.....L}. The first group harbours 
an evolved O or B-type supergiant. These systems typically have a short orbital period \citep{2015A&ARv..23....2W} 
indicating that the supergiant might be close to filling its Roche-lobe. Both mass-transfer modes, 
wind mass-transfer and RLO, could explain the persistent and luminous X-ray emission.
The binaries of the second, more numerous subgroup consist of an early Be type donor and a 
compact object. The appearance of emission lines points to the existence of a circumstellar 
disc of material, thought to be streaming off the nearly critically rotating B\,star along its equatorial plane. Orbiting its host, 
the compact object eventually penetrates this disc and accretes matter at this stage \citep{1985ApJ...292..257A}. 
This process is seen as a transient X-ray source recurring within a few $10$ to $100$ days, 
corresponding to the period of the wide and eccentric orbit. This inefficient accretion mode does not 
affect the orbital separation very much. Thus, the expected X-ray lifetime is the main-sequence time of a B\,star, 
roughly a dozen million years. Building on this, \cite{1989A&A...226...88M} estimated the total numbers of 
supergiant- and Be-star hosting X-ray binaries in our galaxy. Taking observational biases into account, 
they expected about 30 supergiant (SGXB) and 3000 Be-star (BeXBs) hosting X-ray binaries in the Milky Way. 
The fraction of SGXBs to BeXBs reflects the ratio between their expected X-ray life times, i.e. thermal 
to nuclear timescale, which is roughly 1/100. This estimate was in good agreement with observations 
at the time \citep{1995xrbi.nasa.....L}.

\cite{2007yCat..21700175B} \citep[see also][]{2010ApJS..186....1B} discovered that some SGXBs show peculiar behaviour.
While some of them show X-ray transients on a timescale of a few hours \citep{2006csxs.book..267H},
a second group has a characteristic high absorption corresponding to a column density of
$N_\text{H} \geq 10^{23} \,\text{cm}^{-2}$ \citep{2012A&A...547A..20M}. One refers to the first subgroup of
SGXBs as supergiant fast X-ray transients (SFXTs). The latter are called obscured SGXBs.
The mechanisms leading to their formation as well as there role in SGXB evolution are poorly understood.
Considerations of the configuration of the obscured SGXBs reach from the existence of a cocoon of dust enshrouding
the whole system \citep{2008A&A...484..783C} to an unusual slow and dense stellar wind
($v_\infty\sim 400 \,\text{km/s}$) \citep{2011arXiv1105.1988M}. While a dust cocoon could indeed form
due to a common envelope evolution, the reasons why a supergiant donor should exhibit a very slow wind velocity
are not obvious, since observations and numerical calculations do not suggest such slow winds for single stars.
It is therefore reasonable that the high attenuation is somehow connected to the existence of a companion,
rather than an intrinsic attribute of a supergiant.

Recently,
\cite{2015A&ARv..23....2W} show 20 new SGXBs and 8 new BeXBs found by INTEGRAL. This leads to a 
total number of 36 SGXBs versus 60 BeXBs known in the Milky Way. Hence, the current observed number 
of supergiant systems appears too high to be explained by thermal timescale RLO of SGXBs. A way to 
address this problem is to postulate wind accretion in SGXBs 
\citep{2014EPJWC..6402001S, 2016A&A...589A.102B}. However population synthesis studies 
by \cite{1995ApJ...440..280D} predict a number ratio of $\text{SGXBs}/\text{BeXBs} \lesssim 0.15$, 
even if wind accretion is assumed to be the major mass-transfer mode.

On the other hand, the investigation of stabilising processes during RLO is a highly debated field 
of research \citep{2015ebss.book..179I, 2009A&A...507..891D, 1997ASPC..121..361B, 1979A&A....71..352S}. 
\cite{1976ApJ...204L..29P} discussed mass-transfer stabilisation due to the rotational slow down of the donor star, 
caused by tidal breaking, and subsequently diminishing centrifugal force in its outer layers. 
Stabilisation by widening of the Roche-lobe due to mass loss by a stellar wind was studied by \cite{1977ApJ...215..276B}.
\cite{1987ApJ...318..794H} investigated mass-transfer on dynamical timescales using semi-analytical models. 
They found that any initial stability due to the rapid, adiabatic expansion of the primary's outer layer 
will switch to an unstable mode, if these layers are super-adiabatic.
\cite{2000ApJ...530L..93T} discovered the possibility of long term stable mass-transfer even if the mass ration exceeds a value of 4. 
More recently, \cite{2017MNRAS.465.2092P} showed that RLO can be stable if the primary is a post-main sequence star (Case B mass transfer) 
that has already expanded through the Hertzsprung gap but has not yet developed a deep convective envelope 
\cite[see also][]{2015MNRAS.449.4415P}.

A similar timescale problem as in the SGXBs may exist in some ULX sources, many of which 
radiate highly above the Eddington accretion limit of a $\sim10\,\text{M}_\sun$ black hole
\citep{1981ApJ...246L..61L, 2017ARA&A..55..303K}.
An interesting case is the ULX NGC\,7793\,P13. 
In this source, X-ray pulses where discovered by \cite{2017MNRAS.466L..48I}, indicating
that the companion is a neutron star. With an estimated donor star mass of
$\sim 20\,\text{M}_\sun$ 
the mass ratio of the system is large, such that, again, stable mass transfer is not expected to occur. 
However, wind accretion is insufficient to explain the high X-ray luminosity.
How NGC 7793 P13 and three more X-ray pulsating ULXs \citep{2014Natur.514..202B,2017Sci...355..817I, 2018arXiv181104807M} form and transfer mass is therefore not well understood.

In this study, we work out conditions under which stable, nuclear timescale mass transfer can occur
{\em despite} the donor star being much more massive than the compact companion. We explain 
our methods to model single star and binary evolution in Sect.\,2, including mass transfer to a 
compact object, and work out the criteria for long term RLO and their connection to the 
internal structure of the donor. In Sect.\,3 we investigate the sensitivity of the donor star radius 
to mass loss, and we present our binary evolution models in Sect.\,4, including examples of high mass ratio
systems undergoing nuclear timescale mass transfer. In Sect.\,5, we discuss the possible properties of neutron star
and black hole hosting ULXs in the light of our findings. In Sect.\,6 we investigate possible paths
for the evolution of SGXB and ULX progenitors, and in Sect.\,7 we discuss their likely fates, before giving our conclusions in Sect.\,8.

\section{Method}\label{sec:method}
\subsection{Modelling of stellar evolution and mass transfer}
We used the Binary Evolution Code (BEC), a one-dimensional hydrodynamic Lagrangian code to solve the equations of stellar structure \citep{1990sse..book.....K} and model the binary interaction \citep{1993SSRv...66..401B}. The code includes up to date physics \citep{1999A&A...350..148W,2001A&A...369..939W}  and uses the OPAL opacity tables \citep{1996ApJ...464..943I}. For convection zones the mixing length theory (MLT) by \cite{1958ZA.....46..108B} is applied, where we adopted $\alpha_\text{ML}=1.0$ unless stated otherwise. We note that the value of $\alpha_\text{ML}$ is uncertain \citep{2013MNRAS.433.2893P} and may differ for stars in different evolutionary stages and/or mass ranges. Nevertheless, a mixing length parameter of the order of unity is in agreement with observations \citep{2006ApJ...642..225F, 1968pss..book.....C}. 
The rotation and magnetic fields are not taken into account. For stellar wind mass loss we follow the assumptions made by 
\cite{2011A&A...530A.115B}, unless stated otherwise. 

We investigated the case of X-ray binaries, we evolve an ordinary star with a point mass companion, the latter 
representing the neutron star (NS), or black hole (BH). The point mass induces mass transfer when the donor star exceeds its Roche radius, 
and mass is carried to the compact companion accretor via the first Lagrangian point. Here,
the Roche-lobe is approximated by a sphere of radius $R_L$ which has the same volume as the Roche-lobe,
following \cite{1983ApJ...268..368E}
\begin{equation}
\label{roche_radius}
R_L=\frac{0.49q^{2/3}}{0.6q^{2/3}+\ln(1+q^{1/3})}a\, ,
\end{equation}
where $q=M_\mathrm{D}/M_\mathrm{A}$ is the mass ratio of donor and accretor, and  $a$ is the orbital separation. 
The mass-transfer rate is calculated using the method of \cite{1990A&A...236..385K}. 
When the donor star does not fill its Roche volume, we compute the accretion of stellar wind 
material onto the compact star using the description of \cite{1944MNRAS.104..273B}.

When matter falls onto the compact object, it heats up and releases a large fraction of the gained 
gravitational energy in X-rays. This leads to a feedback on the remaining material, of which the radiative force 
may expel a certain fraction. Hence the accretion rate and therefore the X-ray luminosity are self regulated 
and are usually not expected to exceed the Eddington-accretion rate and luminosity of this object (although, see Sect.\,4.6). 
We calculated the Eddington-accretion rate as
\begin{equation}
\label{eddingtonaccretion}
\dot{M}_\text{Edd}=
\begin{cases}
     \num{4.6E-8}\frac{1}{1+X}(M_\text{A}/\text{M}_\odot)^{-1/3}
     ~\text{M}_\sun\text{yr}^{-1}
     \text{for NS} \\[0.3cm]
     \num{7.3E-8} \frac{1}{1+X}(M_\text{A}/\text{M}_\odot)
     ~\text{M}_\sun\text{yr}^{-1}
     \text{for BH} ,
\end{cases}
\end{equation}  
where we distinguish between NSs and the more massive BHs. In this equation, $X$ is the hydrogen mass fraction of the accreted material. The different scaling of the Eddington-rate with respect to the accretor mass arises from different mass-radius relations of BHs and NSs. While the BH's Schwarzschild radius is proportional to its mass, the NS's radius scales as $M_\text{A}^{-1/3}$ for a perfect Fermi gas. We compute the actual accretion rate as the minimum of the mass-transfer rate and the Eddington-rate given by equation \ref{eddingtonaccretion}.

\subsection{Internal H/He-gradients}

\cite{2018A&A...611A..75S} derived the internal H/He-gradients in the hydrogen-rich Wolf-Rayet stars
of the SMC. They found that these gradients are different by more than a factor of ten for different objects.
As we find below that these gradients can play a key role in stabilising the mass transfer in SGXBs, we 
consider it as a free parameter in our models.

We evolved three stellar models of solar metallicity \citep{2011PhDT.........1B} and initial masses of $\SI{50}{M_\sun}$, $\SI{60}{M_\sun}$ and $\SI{80}{M_\sun}$  to core helium mass fractions of $0.6$, $0.7$ and $0.8$.
Subsequently, we adjusted the helium profile by setting the helium mass fraction above the convective core to 
\begin{equation}
Y(m)=\max\left[Y_\text{core}+\frac{\mathrm{d}Y}{\mathrm{d}m}(m-m_\text{core}),~0.2638 \right]\, ,
\label{heprofile}
\end{equation}
with prescribed fixed values of $\mathrm{d}Y/\mathrm{d}m$.
Here $m$ and $Y$ denote the Lagrangian mass coordinate and the helium abundance, $m_\text{core}$ is the mass of the convective core and $Y_\text{core}$ its helium abundance. We then relaxed the models thermally disregarding changes in their chemical profile due to mixing or burning. 
Then we exposed the models to a constant mass-loss rate of $\SI{E-5}{M_\sun yr^{-1}}$ which is of the order of magnitude of the nuclear timescale mass-transfer rate $M_\text{D} / \tau_\text{nuc,~D}$, to the response of the stellar radius and derive the mass-radius exponent
\begin{equation}
\zeta_\text{R}(M):=\frac{\mathrm{d} \ln R}{\mathrm{d} \ln M}\,.
\label{respr}
\end{equation}
The results of this exercise is presented in Sect.\,3.1.

\subsection{Response of the Roche-lobe radius}
\label{srr}

If mass is transferred from the donor to the accretor, the Roche-lobe radius $R_L$ changes due to changing orbital separation. We consider the orbital evolution by changes of the stellar masses and hence of the mass ratio, by angular momentum loss through the donor's stellar wind, and by angular momentum loss by isotropic re-emission of matter near the compact object.
Further effects, such as spin-up of the compact companion, spin-orbit coupling due to tidal effects, magnetic breaking and gravitational wave radiation are neglected.

Supposing a circular orbit, the orbital angular momentum can be written as
\begin{equation}
J=2\pi a^2 \frac{M_\text{A} M_\text{D}}{P M}\,
\end{equation}
where $a$ is the orbital separation, $M_\text{A}$ and $M_\text{D}$ are the accretor mass and the donor mass, 
$M=M_\text{A}+M_\text{D}$ and $P$ is the orbital period. Using Kepler's third law to replace $P$ and solving for $a$, we find 
\begin{equation}
a=\frac{MJ^2}{GM_\text{A}^2M_\text{D}^2}\,.
\end{equation}
The derivative with respect to time provides
the change of orbital separation with time as
\begin{equation}
\frac{\dot{a}}{a}=2\frac{\dot{J}}{J}-2\frac{\dot{M_\mathrm{D}}}{M_\mathrm{D}}-2\frac{\dot{M_\mathrm{A}}}{M_\mathrm{A}}+\frac{\dot{M_\mathrm{D}}+\dot{M_\mathrm{A}}}{M_\mathrm{D}+M_\mathrm{A}}\,.
\label{orbit0}
\end{equation}
Following \cite{2006csxs.book..623T}, we consider the donor's mass to decrease by $\text{d} M_\text{D}$ per time unit. 
Since only stellar wind and mass transfer can change the orbital separation, we introduce $\alpha$ as the 
fraction of $\text{d} M_\text{D}$ lost in a stellar wind and $\beta$ as the fraction that is transferred to the accretor and then re-emitted isotropically with the accretor's specific orbital angular momentum. Hence, we have that the accretion efficiency (accreted mass fraction of $\text{d} M_\text{D}$) $\epsilon=1-\alpha-\beta$. Using this nomenclature, we express the loss of orbital angular momentum as
\begin{equation}
\frac{\dot{J}}{J}=\frac{\alpha+\beta q^2}{1+q} \frac{\dot{M_\mathrm{D}}}{M_\mathrm{D}}\,.
\label{orbit0a}
\end{equation}
Inserting this into Equ.\,(\ref{orbit0}), with $M_\mathrm{A}=M_\mathrm{D}/q$ and $\dot{M_\mathrm{A}}=-\epsilon \dot{M_\mathrm{D}}$, yields after integration
\begin{equation}
\frac{a}{a_0}=\frac{q_0+1}{q+1} \left(\frac{q}{q_0}\right)^{2(\alpha-1)} \left(
\frac{\epsilon q +1}{\epsilon q_0+1}\right)^{3+2(\alpha \epsilon^2 +\beta)/(\epsilon(1-\epsilon))}\, ,
\label{orbit}
\end{equation} 
where the subscript $0$ refers to the initial state before the mass loss.
Investigation of the limits of Equation (\ref{orbit}) using l'Hospital's rule with respect to $\alpha$, $\beta$ and $\epsilon$ helps to understand the orbital behaviour in the extreme cases. For pure wind mass loss ($\alpha=1$), it is
\begin{equation}
\frac{a}{a_0}=\frac{q_0+1}{q+1}\, ,
\label{orbita}
\end{equation}
for dominating mass transfer and re-emission ($\beta=1$)
\begin{equation}
\frac{a}{a_0}=\frac{q_0+1}{q+1} \left(\frac{q_0}{q}\right)^{2} \exp \left[2(q-q_0)\right]\, ,
\label{orbitb}
\end{equation} 
and for conservative mass transfer ($\epsilon=1$) we obtain
\begin{equation}
\frac{a}{a_0}=\left(\frac{q+1}{q_0+1}\right)^4 \left(\frac{q_0}{q}\right)^{2} \, .
\label{orbitc}
\end{equation} 


In HMXBs, all of the three cases tend to decrease the mass ratio, hence $q<q_0$. Keeping this in mind, we find that the orbit will always widen in the case of wind dominated mass loss according to Equation (\ref{orbita}). For mass transfer with subsequent re-emission and accretion, respectively, we find a decreasing orbital separation, since the exponential term in equation 
(\ref{orbitb}) and the fourth power term in equation (\ref{orbitc}) dominate the total change of $a$.

Since massive donor stars have a strong stellar wind, wind mass transfer just before RLO is unavoidable. In this pre-RLO phase, where the accretion is purely wind fed, orbital changes occur due to mass loss and angular momentum loss of the donor star and mass accretion onto the compact object.
It is interesting to study the influence of pure wind accretion on the binary system, since the wind mass loss is the main reason of orbital change in the pre-RLO phase. The widening of the orbit increases the Roche radius. However, a main-sequence donor does not increase its radius much for most of its lifetime. This would inhibit RLO until the donor star is relatively evolved.

If we consider pure wind mass loss and subsequent accretion of a mass fraction $\epsilon$ we can rewrite equation (\ref{orbit0}) and (\ref{orbit0a}) as
\begin{equation}
\frac{\dot{a}}{a}=\left[\frac{2\alpha}{1+q}+2\epsilon q -2 +\frac{\alpha}{1+q^{-1}}\right]\frac{\dot{M_\text{D}}}{M_\text{D}}\, .
\end{equation}
We note that $\dot{M_\text{D}}$ is negative, hence the orbital separation will only decrease if 
\begin{equation}
\frac{2\alpha}{1+q}+2\epsilon q -2 +\frac{\alpha}{1+q^{-1}}>0\,.
\end{equation}
This gives us the condition for a decreasing orbital separation in a phase of pure wind accretion as

\begin{equation}
\frac{\alpha}{\epsilon}<\frac{2(q^2+1)}{q}\,.
\end{equation}
On the other hand, the Bondi-Hoyle-formula can be expressed in terms of orbital and wind velocity
\begin{equation}
\epsilon=\frac{\dot{M}_\text{acc}}{\dot{M}_\text{wind}}=\left(\frac{v_\text{orb}}{v_\text{wind}}\right)^4
\end{equation}
and hence the condition for a shrinking orbit is
\begin{equation}
\left(\frac{v_\text{wind}}{v_\text{orb}}\right)^4-1<\frac{2(q^2+1)}{q}\,.
\end{equation}
Since $\left(v_\text{wind}/v_\text{orb}\right)^4 \gg 1$ and $q^2\gg 1$ we simplify the criterion to
\begin{equation}
\frac{v_\text{wind}}{v_\text{orb}}<\sqrt[4]{2 q}\,.
\end{equation}
For an initially well detached binary system hosting a supergiant donor, the orbit is quite likely to widen, since the orbital velocity will barely exceed a few $100\,\text{km/s}$, while the wind velocity can be an order of magnitude larger. It is hence difficult to start RLO during the early main-sequence phase of the donor star. The stellar wind widens the orbit while the donor's radius hardly increases. A faster expansion, e.g. in the advanced stage of core hydrogen burning ($Y_\text{core}\sim 0.8$), could however overcome the orbital increase induced by the wind mass loss and start a RLO phase. We conclude that donor stars in RLO systems are therefore likely to be evolved.

\begin{figure}[htbp]
\centering
\includegraphics[width=88mm]{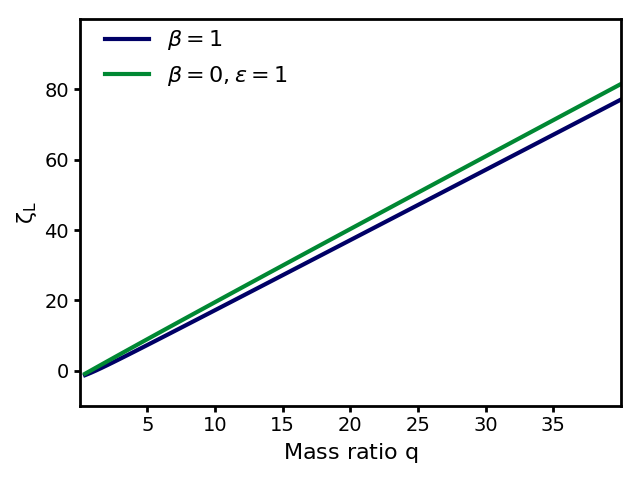}
\caption{Mass-radius exponent of the Roche radius (Equation \ref{resprL1}) as function of mass ratio $q$. The different colors denote different mass loss modes. For the blue line all the mass is transferred and re-emitted isotropically while for the green line all the transferred mass is accreted.}
 \label{response_rl}
\end{figure}

We describe the change of the Roche-lobe radius defining a mass-radius exponent similar to Equation (\ref{respr}) as 
\begin{equation}
\zeta_\text{L}(M):=\frac{\mathrm{d}\ln R_\text{L}}{\mathrm{d} \ln M}\,.
\label{resprL}
\end{equation}
If we use the above-mentioned description of the orbital evolution, we find an analytical expression for the Roche-lobes response to mass transfer, as a function of mass ratio \citep{2006csxs.book..623T} as
\begin{equation}
\zeta_\text{L}(q)=\left[1+\left(1-\beta\right)q\right]\Psi+\left(5-3\beta\right)q\,
\label{resprL1}
\end{equation}
where
\begin{equation}
\Psi=-\frac{4}{3}-\frac{q}{1+q}-\frac{0.4+1/3q^{-1/3}\left(1+q^{1/3}\right)^{-1}}{0.6+q^{-2/3}\ln\left(1+q^{1/3}\right)}\,.
\end{equation}
Figure \ref{response_rl} shows the mass-radius exponent of the Roche-lobe $\zeta_L$ plotted as function of the mass ratio $q$. Since the typical mass ratio of HMXBs is $\geq 8$ we would need  $\zeta_L\geq 12$ in order to find long term mass transfer via RLO.

The evolution of a SGXB where the donor star fills its Roche-lobe depends on the mass-radius exponents. 
If $\zeta_\text{R}\leq \zeta_\text{L}$, the Roche-lobe shrinks faster than the star, 
which will overfill its Roche-lobe even more leading to a higher mass-transfer rate, and so on. 
This runaway process results in a CE evolution. 
If however $\zeta_\text{R}> \zeta_\text{L}$, the donor radius is more sensitive to mass loss than the Roche radius, 
and the mass-transfer rate becomes self-regulated, 
and can be estimated using the 
donor mass and its nuclear timescale as $\dot{M}\approx M_\text{D}/\tau_\text{nuc}$
\citep{1997A&A...327..620S}.

\subsection{Binary evolution models}
After investigating the mass-radius exponent of our single star models, we selected those with high values of $\zeta_\text{R}$.
We added a point mass companions with $2$ or $10 \, \text{M}_\sun$ and started the binary evolution.
The initial orbital separation was selected such that the Roche-lobe radius exceeded the initial photospheric radius 
of the donor by 3\,\%. We used the method mentioned above to calculate the mass transfer and the accretion rate. 
We evolved the models with BEC until hydrogen in the core of the donor star was exhausted or the  
mass-transfer rates exceeded $\sim 10^{-2}\,\text{M}_\sun \text{yr}^{-1}$, where we assume that the mass transfer becomes 
dynamically unstable and the system goes into a common envelope phase. Subsequently, we calculated the X-ray lifetime defined 
as the time interval where the compact companion accretes at a higher rate than 
$10^{-13}\,\text{M}_\sun \text{yr}^{-1}$, which corresponds to an X-ray luminosity of  
$\sim 10^{33}\,\text{erg\,s}^{-1}$.

\section{Results from single star models}\label{sec:results}

Here, we explore how the radii of our potential donor stars are affected by nuclear timescale mass loss.
The resulting values of $\zeta_\text{R}$ can then be compared to the functions $\zeta_\text{L}$ plotted in 
Fig.\,\ref{response_rl} to obtain an estimate for which mass ratios we can expect stable mass transfer.

\begin{figure}[htbp]
\centering
\includegraphics[width=88mm]{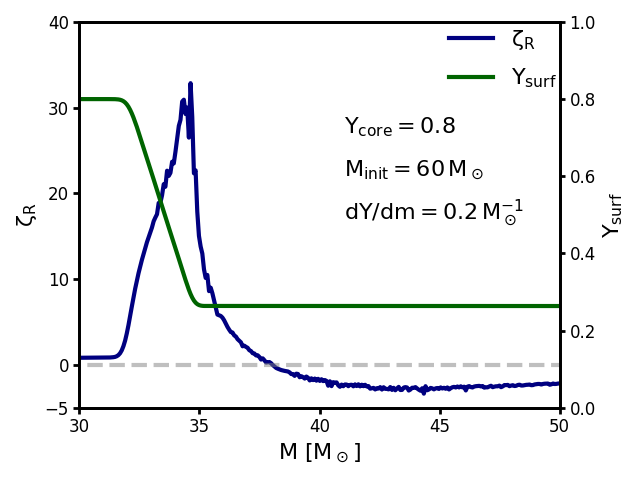}
\caption{Mass-radius exponent (blue) 
of our initial $60\,\mathrm{M}_\sun$ model evolved to a central helium abundance of 0.8, and a H/He-gradient of $0.2\mso^{-1}$ which is
then exposed to a constant mass loss rate of $10^{-5}\msoy$, as function of the remaining
stellar mass. The green line gives the surface helium abundance. 
Since mixing is inhibited during the mass-loss phase, 
the surface helium-abundance evolution reflects the internal helium profile of the initial model.}
\label{m_response_he1}
\end{figure}

Figure \ref{m_response_he1} shows the mass-radius exponent $\zeta_\text{R}$ of our $60\,\mathrm{M}_\sun$ model evolved
to a central helium mass fraction of $Y_{\rm c}=0.8$. At the start of our mass-loss experiment, the model
has a mass of $51\,\mathrm{M}_\sun$, a convective core of $\sim 32\,\mathrm{M}_\sun$, and a H/He-gradient of about
$0.2\,\mathrm{M}_\sun^{-1}$ in the core-envelope transition zone. 

We stripped the mass via a constant mass-loss rate of $10^{-5} \, \text{M}_\sun \text{yr}^{-1}$. 
As mentioned above, this rate is sufficiently small to maintain thermal equilibrium inside the model. 
Figure \ref{m_response_he1} shows that
with decreasing stellar mass, the mass-radius exponent is small and negative at first, due to an increase in the
stellar luminosity-to-mass ratio.
However, when He-enriched layers are getting close to the surface, the mass-radius exponent
climbs to values of 30, before it drops again to a small value when the stellar core is exposed. 
A value of $\zeta_\text{D}=30$ implies that a mass decrease by 1\% induced the radius of the star to decrease by 30\%. 
According to Fig.\,\ref{response_rl}, such high values of $\zeta_\text{D}$ could give rise
to a phase of stable mass transfer even for mass ratios as high as 15. 

The drastic shrinkage of our models is related to the transition from a hydrogen-rich supergiant stage,
with a radius of about $68\,\mathrm{R}_\sun$ to a much more compact and hydrogen-poor Wolf-Rayet type structure with $6\,\mathrm{R}_\sun$.
This is eminent from the correlation of the mass-radius exponent with the change of the surface helium abundance 
as shown in Fig.\,\ref{m_response_he1}. 
Here, we note that the mass-radius exponent changes sign and evolves to large values somewhat {\em before}
helium-enriched layers reach the surface of the star, since it is the average envelop properties which determine its radius.
This is in agreement with previous stellar structure and evolution calculations \citep[e.g.][]{2015A&A...573A..71K,2018A&A...611A..75S}.

\begin{figure}[htbp]
\centering
\includegraphics[width=88mm]{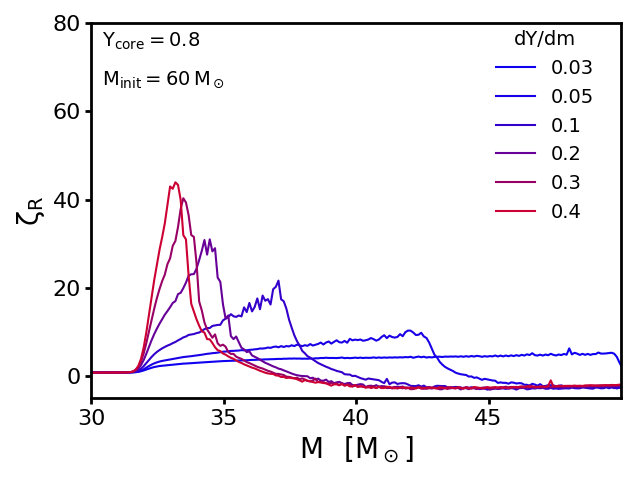}
\caption{ Mass-radius exponent of our initial $60\,\mathrm{M}_\sun$  model as the blue line in Fig.\,\ref{m_response_he1},
but here for models where the internal helium gradient $\mathrm{d}Y/\mathrm{d}m$ has been artificially adjusted (see Sect.\,2) 
to values indicated in the legend.
}
\label{ex60}
\end{figure}

Figure\,\ref{ex60} shows that the internal H/He-gradient is a suitable way to tune the
the mass-radius exponent $\zeta_\text{D}$ in our models. It depicts the result of the same experiment
explained above, but for six models with different steepnesses of the H/He-gradient. 
It can be seen that the models with steeper gradients reach higher values of $\zeta_\text{D}$, 
even exceeding $\zeta_\text{D}=40$ in the most extreme case. While it may appear surprising at first, since such high
values of the mass-radius exponent have not yet been reported in the literature, it is a simple consequence
of the mass in the transition layer between the He-rich core and the H-rich envelope, $\Delta M_{\rm H/He}$, 
becoming very small for a steep internal H/He-gradient, and 
$\zeta_\text{D} \simeq (R_{\rm H}-R_{\rm He}) M / (R_{\rm He} \Delta M_{\rm H/He})$ becoming larger the smaller $\Delta M_{\rm H/He} \rightarrow 0$.  
Here, $R_{\rm H}$ is the stellar radius in the H-rich state, and $R_{\rm He}$ the one in the He-rich state, while
$M$ is the mass of the star.

Infinite H/He-gradients, though not strictly excluded, are not expected in massive stars. However, it is important to point
out that the range in steepnesses explored in Fig.\,\ref{ex60} remains well within the range which has been derived
by \cite{2018A&A...611A..75S} from the observed properties of the WN-type Wolf-Rayet stars in the SMC. 
As we discuss in Sect.\,5 below,
this group of stars is quite relevant here, since SGXBs may evolve into WN-type Wolf-Rayet binaries. 

\begin{figure}[htbp]
\includegraphics[width=83mm]{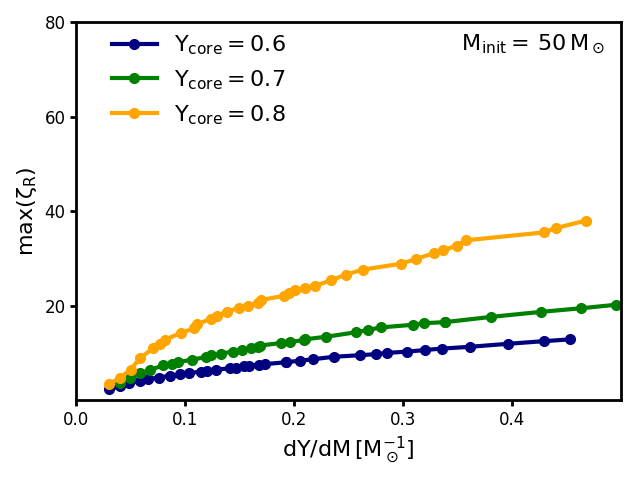}
\includegraphics[width=83mm]{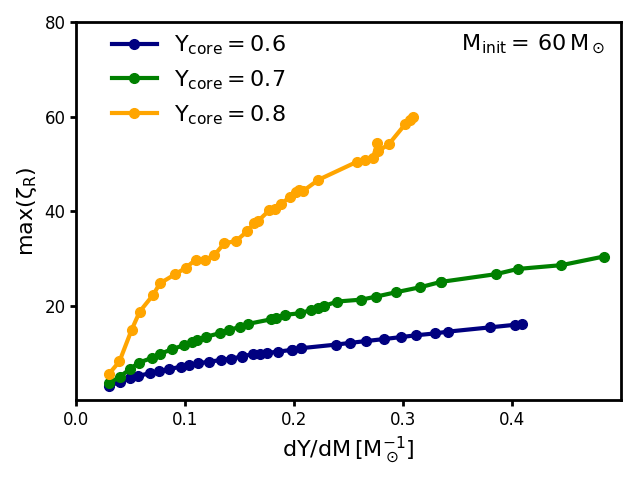}
\includegraphics[width=83mm]{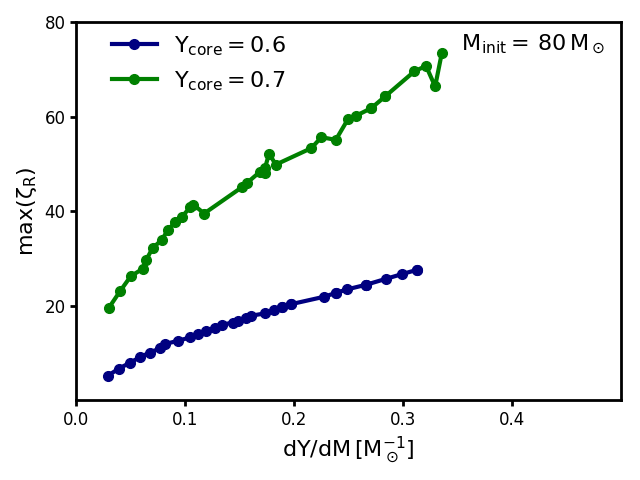}
\caption{Maximum value of the mass-radius exponent $\zeta_\text{R}$ as function of the chosen internal helium gradient $\mathrm{d}Y/\mathrm{d}m$
for stellar models derived from three different initial masses as indicated by the legends. 
Different colours indicate different core helium abundances at the start of the  mass loss experiments (see Sect.\,2). }
\label{maxresp_dydm}
\end{figure}

The mass-radius exponents of all our single star models 
($\SI{50}{M_\sun}$, $\SI{60}{M_\sun}$, $\SI{80}{M_\sun}$) and core helium abundances ($0.6$, $0.7$, $0.8$)
are shown in the Appendix (Figures \ref{respall1} - \ref{respall3}), where we explore six different H/He-gradients
per models, as in Fig.\,\ref{ex60}. These figures show that besides the clear correlation between 
the mass-radius exponent and the helium gradient, larger mass-radius exponents are also obtained for
a higher initial mass, and for a later evolutionary stage (larger core helium mass fraction).
Figure \ref{maxresp_dydm} summarises these results, in showing the maximum value of the mass-radius exponent $\zeta_\text{R}$
as function of the adopted internal helium gradient $\mathrm{d}Y/\mathrm{d}m$ for all our single star models.

\subsection{Influence of stellar inflation}

\begin{figure}[htbp]
\includegraphics[width=88mm]{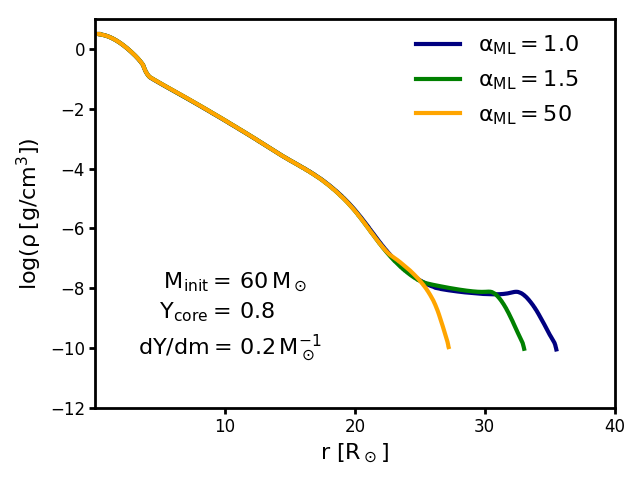}
\caption{Mass density as function of radius for our initially \protect  $60\,\text{M}_\sun$ 
models with $Y_\mathrm{core}=0.8$ and $\mathrm{d}Y/\mathrm{d}m=0.2\,\text{M}_\sun^{-1}$, 
for three different values of the mixing length parameter $\alpha_\mathrm{ML}$, as indicated. }
\label{resp_infl1}
\end{figure}

In their analysis of the massive single star evolutionary models for LMC composition of \cite{2011A&A...530A.115B}, \cite{2015A&A...573A..71K} and \cite{2015A&A...580A..20S} found that the envelopes of models for stars
above $\sim 40 \, \text{M}_\sun$ are inflated, since they exceed the Eddington-limit in their subsurface layers.
Whether such inflated envelopes (see Fig.\,\ref{resp_infl1} for examples) exist in reality is still a matter of debate,
although they have been confirmed by 3D-radiation-hydrodynamic calculations \citep{2015PASJ...67..118J}.

The mass of the inflated envelope is mostly very small, i.e., about $10^{-6}\,\text{M}_\sun$ in our models. 
However, its radius may be of the order of the radius of the un-inflated stellar interior. Obviously, radius inflation
may play a big role in binary evolution.
It is hence important to analyse how inflation may affect the mass-radius exponent. 

For this purpose, we investigated our $60\,\mathrm{M}_\sun$ single star model at a core helium mass fraction
of $Y_\mathrm{core}=0.8$ and $\mathrm{d}Y/\mathrm{d}m=0.2\,\text{M}_\sun^{-1}$, from which we constructed three different initial models for our mass-loss experiment.
The only difference in these models is the chosen mixing length parameter $\alpha_\mathrm{ML}=l/H_P$, 
were $l$ denotes the mixing length and $H_P$ the pressure scale height (cf., Sect.\,2).
We computed models with values for $\alpha_\mathrm{ML}$ of 1 (our default choice), 1.5 and 50.
Whereas the first two values are in the range discussed in comparison to 3D-models of convection
and real stars \citep{2019A&A...621A..84S}, we use $\alpha_\mathrm{ML}=50$ to produce a stellar
model in which inflation is suppressed, even though not absent \citep[cf.][] {2015A&A...580A..20S}.

\begin{figure}[htbp!]
\includegraphics[width=88mm]{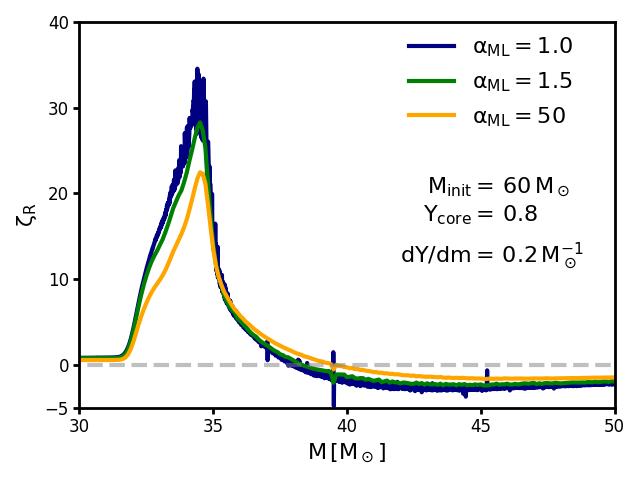}
\caption{\label{resp_infl2} Mass-radius exponent as function of stellar mass, with the three models displayed in
Fig.\,\ref{resp_infl1} as initial models before assuming a constant mass loss rate of $10^{-5}\msoy$.
The colours correspond to the density profiles in Figure \ref{resp_infl1}.}
\end{figure}

The mass-radius exponents of these three models is plotted in Fig.\,\ref{resp_infl2}. We find that stellar inflation 
has a significant effect on $\zeta_\text{R}$. While for strong inflation ($\alpha_\mathrm{ML}=1$) the maximum value 
of $\zeta_\text{R}$ reaches 34, it only reaches 21 if inflation is suppressed ($\alpha_\mathrm{ML}=50$). 
We also note that initially the value of the mass-radius exponent is more negative for the inflated models. 
However the high peak of $\zeta_\text{R}$ occurs only when a H/He-gradient appears beneath the surface.

\begin{figure}[htbp!]
\includegraphics[width=88mm]{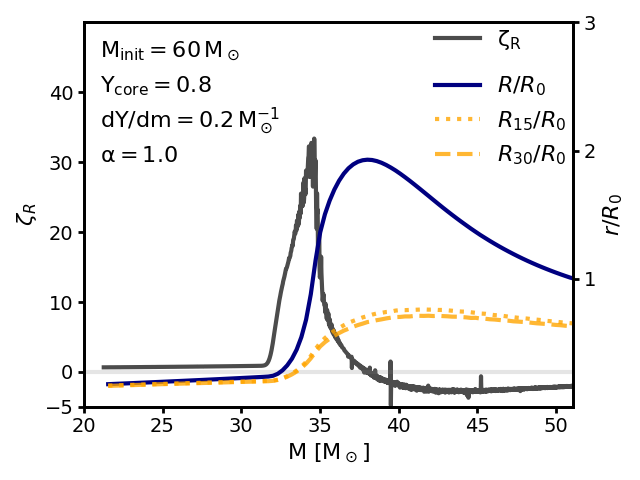}
\includegraphics[width=88mm]{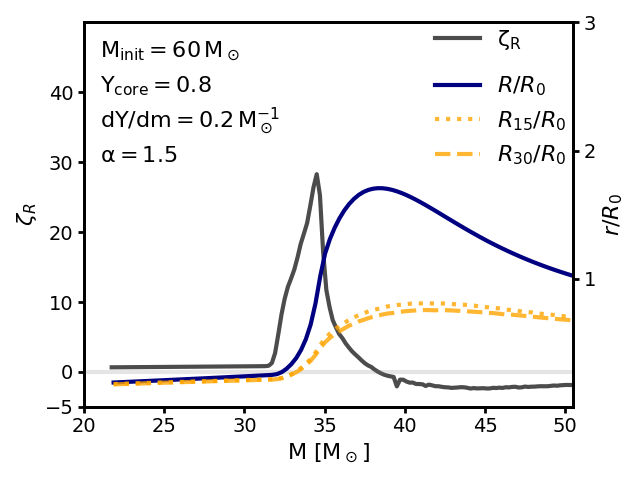}
\includegraphics[width=88mm]{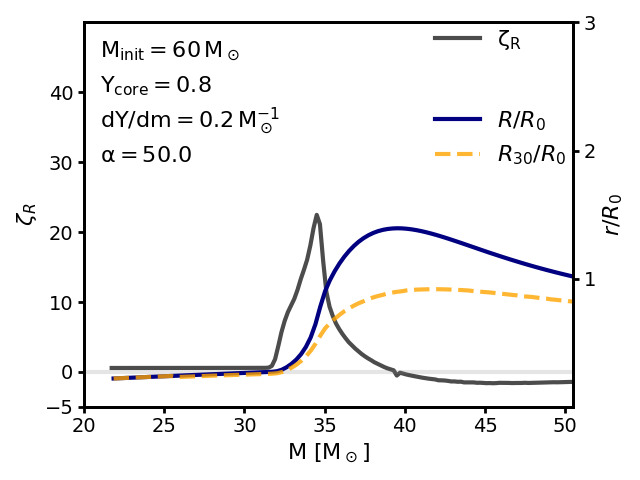}
\caption{\label{resp_infl3} Mass-radius exponent $\zeta_{\rm R}$ (grey), stellar radius $R$ (blue), and radii 
where the gas pressure contributes only 15\% ($R_{15}$) and 30\% ($R_{30}$) to the total pressure (see text),
as function of the remaining stellar mass, for the three models with different values of the mixing length parameter
$\alpha_{ML}$ shown in Fig.\,\ref{resp_infl2} (the lines for $\zeta_{\rm R}$ are identical). All radii are measured
in units of the initial stellar radii $R_0$, which can be read off Fig.\,\ref{resp_infl1}.}
\end{figure}

In Fig.\,\ref{resp_infl3}, we compare the radius extension of the un-inflated part with that
of the inflated envelope layer during the mass loss experiment for our three models. 
Here, we follow \cite{2015A&A...580A..20S} to define the bottom of the inflated envelope
as the point, where gas pressure contribution $P_\text{gas}/(P_\text{gas}+P_\text{rad})$
drops to 15\% for the first time when going from the stellar center outwards. 
We refer to this position as $R_{15}$. As the gas pressure fraction in our model computed
with $\alpha_\mathrm{ML}=50$ never drops below 15\%, we define the radius $R_{30}$ 
as the position where the gas pressure fraction first drops below 30\%. The near coincidence of
both radii in the top and middle panel of Fig.\,\ref{resp_infl3} suggest that the exact threshold
value of the gas pressure fraction in defining the bottom of the inflated envelop is not important.

Figure \ref{resp_infl3} shows that before the helium gradient appears beneath the stellar surface
(cf., Fig.\,\ref{m_response_he1}), the stellar radius expands significantly due to inflation. This effect is stronger
in the model computed with lower $\alpha_\mathrm{ML}$, as the radius of the un-inflated part of the star
(defined through $R_{15}$ and $R_{30}$ in Fig.\,\ref{resp_infl3}) behaves the same in all three cases.

Since all three models end up in the same configuration once they are stripped down to the helium core,
since inflation is a much smaller effect for hot and compact models \citep{2015A&A...580A..20S},
Fig.\,\ref{resp_infl3} offers a simple explanation of the dependence of the maximum of the mass-radius 
exponent on the mixing length parameter found in Fig.\,\ref{resp_infl2}). For smaller $\alpha_\mathrm{ML}$,
the hydrogen-rich models are more extended, and the drop in radius towards the compact stage is thus stronger,
compared to when $\alpha_\mathrm{ML}$ is larger. We conclude that inflation is not a critical factor in producing
large mass-radius exponents, but that it can contribute at the quantitative level, i.e., enhancing the
mass-radius exponent by factors of the order of\,2 for stars which exceed the Eddington limit.


\section{Binary evolution models for SGXBs}

In the previous section, we found that models of supergiant stars may show very large
mass-radius exponents, with values up to $\sim 40$. Comparing those to the 
mass-radius exponents of the Roche-lobe radius in Fig.\,\ref{response_rl} leads to the 
expectation that nuclear timescale mass transfer may occur even in binaries with mass ratios of 20 or more. 
To demonstrate this, we construct appropriate initial models and combine them with point masses
in model binary systems, and perform detailed binary evolution calculations
of such systems with our binary evolution code (BEC).

\begin{figure}[htbp!]
\includegraphics[width=88mm]{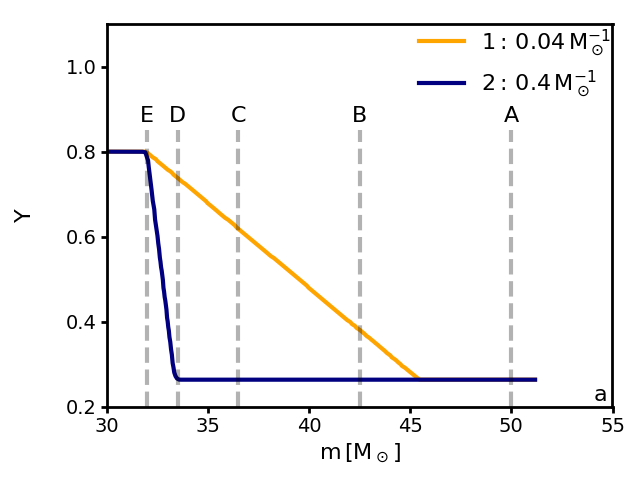} 
\includegraphics[width=88mm]{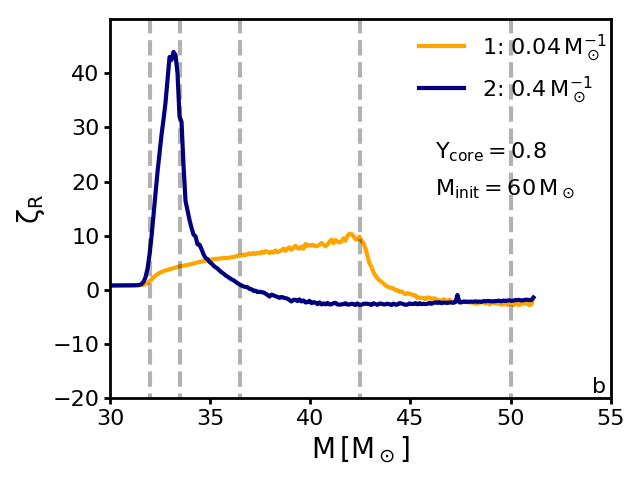}
\caption{Top panel: Helium abundance as function of the mass coordinate of the two stellar models from which the initial donor star models
for our binary evolution calculations are derived. They are constructed such that mass located above 
the lines labelled A\,to\,E was removed before the binary calculation was started.
Bottom panel: Mass-radius exponent for the models, as function of the remaining stellar mass. 
Dashed lines correspond to the masses (Labels\,A to\,E) in the upper panel.}
\label{bin_evo_mr}
\end{figure}

We draw our initial models for these calculations from our $60\,\mathrm{M}_\sun$ models with a central helium mass fraction 
of $Y_\text{core}=0.8$, from which we took one model with a rather shallow helium gradient ($\mathrm{d}Y/\mathrm{d}m=0.04\,\text{M}^{-1}_\sun$)
and a second one with a 10-times steeper helium gradient ($\mathrm{d}Y/\mathrm{d}m=0.4\,\text{M}^{-1}_\sun$). The helium profiles
of these two models are shown in Fig.\,\ref{bin_evo_mr}.
We note that a helium gradient of $\mathrm{d}Y/\mathrm{d}m=0.04\,\text{M}^{-1}_\sun$ corresponds to the gradient which is
left by the retreating convective core during core hydrogen burning, whereas an about 10-times steeper gradient
can be established above the helium core during hydrogen shell burning, as derived for the SMC WR stars by
\cite{2018A&A...611A..75S}. 

From each of those two models, we constructed five different initial models for the binary evolution calculations,
by removing the envelope mass down to the mass indicated by the Labels\,A$\dots$E in the top panel of Fig.\,\ref{bin_evo_mr}. 
The choice of these amounts of removed envelope mass is becoming clear from the bottom panel of Fig.\,\ref{bin_evo_mr},
which shows the mass-radius exponent of both models as function of the remaining mass. 
It indicates that with the chosen envelope masses, our binary evolution models sample the possible range of the
{\em initial} mass-radius exponent of the donor star. 

For each of the 10\,initial donor star models described above, we performed several binary evolution calculations.
We considered two different compact objects, a $2\,\mathrm{M}_\sun$ neutron star and a $2\,\mathrm{M}_\sun$ black hole. Furthermore,
we ran models with different assumptions on the donor star's stellar wind-mass loss, i.e., without wind,
with a constant stellar wind-mass loss rate, and with the mass loss rate according to the prescription of \cite{2001A&A...369..574V}.
Table\,\ref{tab} gives an overview of the different binary evolution models, the details of the initial donor star models,
and key quantities describing the evolution of the model binaries.

For the following discussion, we label the initial donor star models depending on their initial mass (using the letter
A\,to\,E according to Fig.\,\ref{bin_evo_mr}), and the numbers\,1 or\,2 depending on their helium gradient.
For instance, the Model\,A1 refers to the initial donor star model with a helium slope of $\mathrm{d}Y/\mathrm{d}m=0.04\,\text{M}^{-1}_\sun$
(orange curve in Fig.\,\ref{bin_evo_mr}) and an initial mass of $50\,\text{M}_\sun$, while Model\,E2 has $\mathrm{d}Y/\mathrm{d}m=0.4\,\text{M}^{-1}_\sun$ 
(blue curve in Fig.\,\ref{bin_evo_mr}) and an initial mass of  $32\,\text{M}_\sun$.

\subsection{A detailed example: Model\,D2 with a neutron star companion}
\label{modeld2}

Here, we discuss one of our binary evolution models in detail. For this we chose Model\,D2 as donor star, put together with
a $2\,\mathrm{M}_\sun$ neutron star in an 9.1\,d orbit. The initial mass ratio in this binary is 16.8, such that highly unstable mass
transfer might be expected. However, the envelope mass of Model\,D2 was chosen such that the steep helium gradient
in the initial donor model is located just beneath the surface, such that we expect an initial mass-radius exponent of
$\zeta_{\rm R} \simeq 40$ in this case (Fig.\,\ref{bin_evo_mr}).  

\begin{figure}[htbp!]
\includegraphics[width=0.49\textwidth]{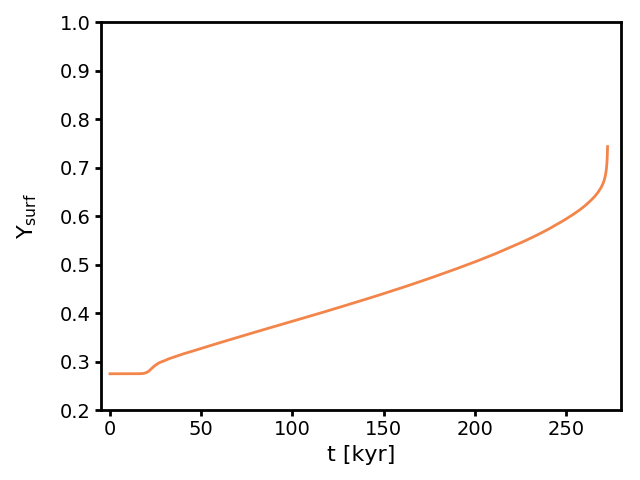}
\includegraphics[width=0.49\textwidth]{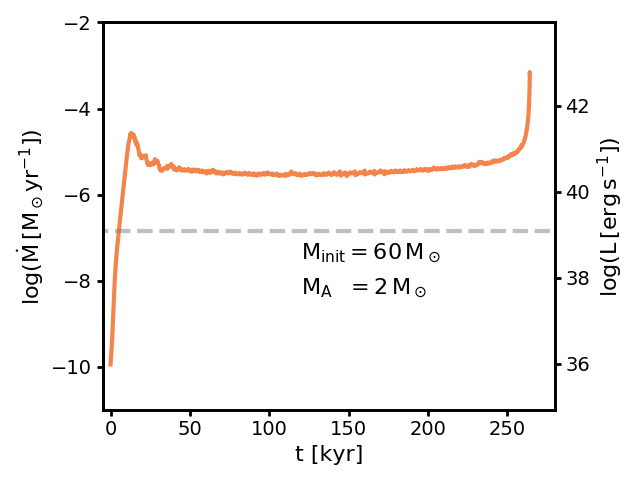}
\caption{Evolutionary properties of our Model\,D2 with a $2\,\mathrm{M}_\sun$ NS companion and stellar winds neglected.
Top panel: surface helium mass fraction of the donor star as function of time during the mass-transfer phase.
Bottom panel:
Evolution of the mass-transfer rate for the same model (left Y-axis).
The right axis indicates the X-ray luminosity corresponding to the mass-transfer rate
(not Eddington limited) assuming an accretion efficiency of $\eta=0.15$. The dashed line gives the Eddington
accretion limit for the neutron star which is applied in this calculation. 
Despite the high initial mass ratio of 16.8, the model settles into
a stable mass transfer for about 0.25\,Myr, with a mass-transfer rate of the order of the nuclear timescale mass-transfer rate.}
\label{D2}
\end{figure}

Figure\,\ref{D2} shows the evolution of the mass-transfer rate as function of time for this model, where stellar 
wind mass loss is neglected. 
It shows that after a brief switch-on phase ($\sim 10^4\,\mathrm{yr}$) the model establishes a rather stationary
mass-transfer rate of $3\, 10^{-6}\,\text{M}_\sun\text{yr}^{-1}$, which is maintained for $\sim 250\,000\,$yr. 
During this time, the orbital period decreased from 9.1\,d to 1.4\,d. This was possible without leading to 
a common envelope situation because the donor star radius shrank from $36\rso$ to $10\rso$ at the same time. 
This shrinking of the donor star was enabled by the continuously increasing surface helium abundance
during the mass transfer (Fig.\,\ref{D2}). I.e., the donor star starts the mass-transfer phase as an early B\,type supergiant ($T_{\rm eff}\simeq 25\,$kK) and ends it as a late-type WNh star ($T_{\rm eff}\simeq 50\,$kK). 

Figure\,\ref{D2} also gives an indication of the X-ray luminosity which might be expected from binaries similar
to our System\,D2. In the evolutionary calculations, we assumed Eddington limited accretion onto the neutron star,
which would produce an X-ray luminosity of the order of $10^{39}\egs$ (dashed horizontal line), comparable to what is found in some
SGXBs. However, Fig.\,\ref{D2} also shows that if the neutron star could accrete at super-Eddington rates,
X-ray of up to $10^{41}\egs$ could be achieved. We discuss this possibility further in Sect.\,5.1.   

The mass transfer lasts for about 0.25 Myr with a mass-transfer rate of $3\times 10^{-6}\,\text{M}_\sun\text{yr}^{-1}$, 
which is two orders of magnitude above the Eddington limit of $3.6\times 10^{-8}\,\text{M}_\sun\text{yr}^{-1}$ of a neutron star, 
and corresponds to an accretion luminosity $2.5\times 10^{40}\,\text{erg}\text{s}^{-1}$ (Fig.,\ref{D2}). 
The orbital period decreases from 9 days to 1 day. Since the mass-transfer rate is much higher than the Eddington accretion limit, 
most of the transferred mass in this calculation is re-emitted. 
According to equation (\ref{orbitb}), the orbital separation shrinks exponentially with the mass ratio. 
Since $q\propto M_\text{D}$ and $\dot{M}_\text{RLO}$ is roughly constant during most of the mass-transfer phase, 
the orbital separation shrinks also exponentially with respect to time.

\begin{figure}[tb!]
\includegraphics[width=0.49\textwidth]{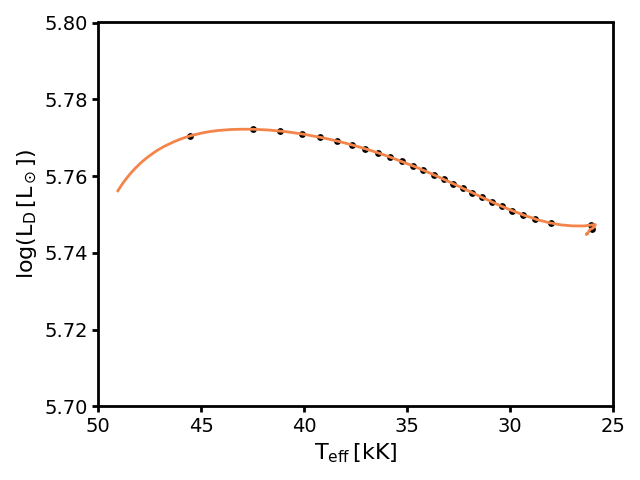}
\caption{Hertzsprung-Russel diagram   for the donor star of the binary model including the initial donor Model D2 and a $2\,\text{M}_\sun$ neutron star accretor. Black dots correspond to time differences of $10^4$ yr.}
\label{hrd_kippenhan1}
\end{figure}

During the mass transfer about $1.5\,\text{M}_\sun$ are removed. 
Figure \ref{hrd_kippenhan1} shows the evolution of the donor star during the mass-transfer phase in the HR diagram. 
The evolutionary track starts at $T_{\rm eff}\simeq 25\,\text{kK}$). As discussed before, 
the donor star becomes hotter and slightly more luminous during the mass-transfer phase. Is temperature remains the longest time
between $25\dots 40\,$kK, which coincides with the regime where SGXB donor stars are observed (cf., Sect.\,6.1). 

The connection of increasing surface helium abundance and increasing effective temperature was already recognized by \cite{2011PhDT.........1B} and \cite{2015A&A...573A..71K} during the evolution of massive single stars. In the cited works, the increasing surface helium abundance was due to rotational mixing and strong wind mass loss. Hence, only massive stellar models ($>60\,\text{M}_\sun$) with high rotational velocities evolved to the hot part of the HR-diagram during core hydrogen burning.

%
%
\subsection{Models without stellar winds}
We first discuss the binary models with a a shallow helium gradient of $0.04\,\text{M}^{-1}_\sun$ neglecting stellar wind mass loss. 
Figure\,\ref{bin_evo_1_bh} shows the evolution of mass-transfer rate, orbital period and donor mass with time for the donor 
Models A1, B1, C1, D1 and E1 described above and a $10\,\text{M}_\sun$ black hole accretor. Of these,
Models B1 and C1 undergo an extended phase of mass transfer of $0.5\,\text{Myr}$ and $0.15\,\text{Myr}$, respectively. 
Despite the rather high mass ratio of $q \simeq 4$, this is much longer than thermal timescale of the donor star ($\sim 10^4 \,\text{yr}$). 

Figure\,\ref{bin_evo_1_bh} also shows the X-ray luminosity corresponding to the mass-transfer rate, assuming
an accretion efficiency of a non-rotating black hole ($\eta =0.06$). Since the  
accretion rate might be Eddington limited, this shows the maximally achievable X-ray luminosity, where
realistic values are expected between this and value corresponding to the Eddington accretion rate for a non-rotating black hole.   
For super-Eddington accretion in these systems, a luminosity up to $\sim 10^{40}\,\text{erg/s}$ could be achieved. 
This matches the order of magnitude of X-ray luminosities observed in ultra luminous X-ray sources. We note that our models do 
not only show a high mass-transfer rate, but also a long X-ray life time.

It is a remarkable feature of both binary models, B1 and C1, that the mass-transfer rate increases steeply after some hundred thousand years. 
This occurs at the time where the flat inner part of the helium profile ($Y(m)=0.8$; cf., Fig.\,\ref{bin_evo_mr}) 
reaches the surface, supporting the idea that a steep H/He-gradient is needed to stabilise the mass transfer. 
As soon as the helium profile close to the surface is flat again, the mass transfer becomes unstable, as expected for a high mass ratio.

Why do only Models B1 and C1 show a long term mass transfer? As shown by Fig.\,\ref{bin_evo_mr}, 
Model\,A1 contains a massive hydrogen rich envelope. 
The companion has to remove about $5\,\text{M}_\sun$ to dig out the H/He-transition layer. 
By doing so the orbit shrinks dramatically (cf., Tab\,\ref{tab}). 
Indeed if we assume that most of the transferred mass is re-emitted, we find according to Eq.\,(\ref{orbitb}) $a/a_0=0.42$. 
This means the orbital separation and hence the donor's radius halve even before the helium gradient scratches the surface.

The mass transfer in Models\,D1 and E1, on the other side, is not long-term stable since the donor star 
has already lost so much mass that the helium-rich plateau is close to or even at the surface at the beginning of mass transfer 
as can be seen in Figure \ref{bin_evo_mr}. The donor star in Model\,D1 needs to lose only  $~1\,\text{M}_\sun$ 
to find the flat helium profile in its outer envelope. 
For a nuclear timescale mass-transfer rate of $2\times 10^{-5}\,\text{M}_\sun$  (Fig.\,\ref{bin_evo_1_bh}), 
it would take only $\sim 50.000$\,yr to remove this layer. This is of the order of the thermal timescale. 
Thus, such a stable mass-transfer can hardly be distinguished from a runaway on a thermal timescale.

The evolution of the Models A1 to E1 are also shown including a neutron star companion in Fig\,\ref{bin_evo_1_ns}. 
We see, that none of these binary models undergoes nuclear timescale mass transfer. This is not surprising since Fig.\,\ref{bin_evo_mr} 
suggests a maximum mass-radius exponent of $\zeta_\text{R}\sim 10$. Since the initial mass ratio is $\sim 20$, 
we find according to Fig,\,\ref{response_rl} $\zeta_\text{L}\sim 30$ and thereby $\zeta_\text{R}<\zeta_\text{L}$. 
Since the Roche radius is more sensitive to mass transfers than the donor radius, a runaway on a thermal timescale is unavoidable. 
Thus, the mass transfer for donor Models A1 to E1 is not sufficiently stabilized by the helium gradient to allow nuclear timescale
evolution.

However, our models with a steeper H/He-gradient can lead to nuclear timescale mass transfer also in the case of a neutron star accretor,
as we have seen in Sect.\,\ref{modeld2}.
Figure\,\ref{bin_evo_1_ns} displays the evolution of the mass- transfer rate for our Models\,A2 to\,E2, 
which differ from Models\,A1 to\,E1 only in the slope of the helium gradient. 
We see that Model\,D2 is the only one which undergoes nuclear timescale mass transfer, due to the more extreme
mass ratio compared to the case of black hole accretors. Still, also for neutron star accretors, nuclear timescale mass transfer
is clearly possible if the donor star's outer envelope possesses a steep H/He-gradient.  

Finally, Fig.\,\ref{bin_evo_1_bh}  shows the mass-transfer rates and corresponding luminosities 
of our binary models composed of a donor with a steep H/He-gradient ($\mathrm{d}Y/\mathrm{d}m=0.4\,\text{M}^{-1}_\sun$) 
(Models\,A2 to\,E2) and a $10\,\text{M}_\sun$ black hole companion. 
Similar to the case of donors with the shallower H/He-gradient, only Models\,C2 and\,D2
develop nuclear timescale mass transfer. However, due to the larger mass-radius exponents
of Models\,C2 and\,D2, the mass-transfer rates remain somewhat smaller. Due to this, and since
the initial radii of Models\,C2 and\,D2 are somewhat larger than those of Models\,C1 and\,D1, the mass 
transfer lasts for about 650\,000\,yr in both cases.

Both donor stars starts to contract towards core helium ignition, such that the mass-transfer rate drops, and
we end our calculations. We consider the further evolution of our systems qualitatively in Sect.\,\ref{fates}.


\subsection{Models including stellar winds}
We showed that our binary models may undergo long-term mass transfer if the H/He-transition layer of the donor star is close to the surface. 
Furthermore, in order to obtain nuclear timescale mass transfer, the H/He-gradient needs to be steeper in the case of neutron star 
accretors compared to that of black hole accretors, due to the more extreme mass ratio in the former.
Here, we assess the question whether an additional mass loss due to a stellar wind mass from the donor star 
could have an additional, perhaps stabilizing effect. 
To investigate this, we performed the same binary calculations as in the previous subsection but with an additional 
constant donor wind mass-loss rate of  $10^{-6}\,\text{M}_\odot/\text{yr}$, 
or, alternatively, with the mass-loss rate as given by \cite{2001A&A...369..574V}.
We restrict these calculations to the donor star models which include the steep H/He-gradient (Models\,A2 to\,D2).

We start our discussion with the binaries hosting BH accretor (Fig.\,\ref{bin_evo_1_bh}). 
Comparing the calculations including the two wind recipes to the ones without any wind, we find that in the case of 
constant mass loss rate, they differ only slightly. The most import difference here is, that the mass-transfer rate 
does not exceed the Eddington accretion limit in the late phase of mass transfer if a wind is included. 
The X-ray life time as well as the orbital separation do not change very much compared to Fig.\,\ref{bin_evo_1_bh}.
This is to be expected, as the wind mass-loss rate of $10^{-6}\,\text{M}_\odot/\text{yr}$ is comparable to the mass-transfer
rate in this case.

The situation is different if we include Vink's mass-loss scheme. The predicted mass-loss rates predicted are higher 
by an order of magnitude compared to the constant mass loss rate discussed before ($\dot{M}_\text{w}\sim10^{-5}\,\text{M}_\odot/\text{yr}$), 
as shown with the dotted line in Fig.\,\ref{bin_evo_1_bh}. 
Vink's wind mass-loss rate in our models is of the same order, or even higher than the nuclear timescale  mass-transfer rate 
for the corresponding binary models inferred without any stellar wind. This means that the donor star can shrink 
only due to its own wind mass loss. Mass transfer does not occur on a nuclear timescale since the Roche-lobe overflow is no longer self-regulated. 
In one case (Model D2), Roche-lobe overflow is not even initiated since the donor avoids any expansion. 
As discussed above, any stellar wind will expand 
the orbital separation. Both the fast shrinking donor radius and the expanding orbital separation drive the systems far away from Roche-lobe filling. 
On the other hand, if the hydrogen-rich envelope is not yet removed or if the former convective core with a flat helium gradient is already at the surface, 
the radius is not sensitive enough to mass loss any more. In theses cases, even a strong stellar wind does not help to keep the system stable.

In the case of NS accretors, we find that a donor wind could have a substantially stabilising effect. This can be seen at the example 
of Model\,D2 in Figure \ref{bin_evo_1_ns}, for which a constant mass loss rate of $10^{-6}\,\text{M}_\odot/\text{yr}$ extends
the mass-transfer phase from $250\,\text{kyr}$ in case of no wind to more than $600\,\text{kyr}$. However, also here larger
stellar wind mass loss rate can have the opposite effect. The Model\,D2 does not undergo Roche-lobe overflow 
when the mass loss rate from Vink is applied.

We conclude that the donor star winds may play an important role in determining the duration of the mass transfer
phase and thus the X-ray lifetime of SGXBs, and which they may extend or decrease. We emphasise that the mass loss rates
of helium-enriched OB\,supergiants are uncertain by more than a factor of two \citep{2017A&A...600A..81R}.
In addition, wind clumping \citep{2018MNRAS.475.3240E} and X-ray emission in SGXBs may affect 
the donor stars wind \citep[e.g.][]{2018A&A...610A..60S}. A more detailed study of the influence of the donor wind
on the SGXB evolution is therefore clearly warranted, but is beyond the scope of our present paper.

\subsection{Orbital period derivatives}
As discussed in Sect.\,\ref{srr}, mass transfer induces changes of the orbital separation. 
These changes may be observed as changes of the orbital period. To show the impact of mass transfer 
due to Roche-lobe overflow on the orbital period, we derive expressions for the orbital period derivative for the case 
of pure isotropic re-emission($\beta=1$), and for conservative mass transfer ($\epsilon=1$). 
Equations\,\ref{orbitb} and\,\ref{orbitc} describe the change of the orbital separation for these cases.

The derivative of Kepler's third law with respect to time leads to
\begin{equation}
\frac{\dot{P}}{P}=\frac{3}{2}\frac{\dot{a}}{a}-\frac{1}{2}\frac{\dot{M}}{M}\,.
\end{equation}
When the change of orbital separation is expressed as $a/a_0=f(q)$, as in  Eqs.\,\ref{orbitb} and\,\ref{orbitc}, then 
\begin{equation}
\frac{\dot{a}}{a}=\frac{\mathrm{d} \ln f}{\mathrm{d}q}\,\dot{q}\,.
\end{equation}
With $\dot{M}=\dot{M}_1$ for the case of isotropic re-emission, we find
\begin{equation}
\label{pdotpa}
\frac{\dot{P}}{P}=\left(3q-2\frac{q}{q+1}-3\right)\frac{\dot{ M}_1}{M_1} ,
\end{equation}
and with $\dot{M}=0$ for conservative mass transfer
\begin{equation}
\label{pdotpb}
\frac{\dot{P}}{P}=3(q-1) \,\frac{\dot{ M}_1}{M_1} .
\end{equation}
For both cases, i.e., from Eqs.\,\ref{pdotpa} and \ref{pdotpb}, a large mass ratio ($q\gg 1$) implies
\begin{equation}
\frac{\dot{P}}{P} \simeq 3q \,\frac{\dot{ M}_1}{M_1} .
\label{pdotp}
\end{equation}

Thus, we expect the orbital period derivative
for Roche-lobe overflow driven mass transfer 
to be essentially independent of the accretion rate of the compact object, as long as the mass ratio is large.
Furthermore, $\dot{M}_1 <0$ implies a decrease of the orbital period during mass transfer, no matter how conservative the mass transfer is. 
Hence, isotropic re-emission and conservative mass transfer lead to an orbital decay at a very similar rate.


The orbital period of model D2, with a donor mass of $\sim 30\, \mathrm{M}_\sun$ (Tab. \ref{tab}) 
and a mass-transfer rate of $\sim 3\times 10^{-6}\, \mathrm{M}_\sun\mathrm{/yr}$ (Fig. \ref{D2}), 
decreases with $\dot{P}/P\approx -4.5\times 10^{-6}\mathrm{yr}^{-1}$, for both, BH or NS companion. 
For donor Models\,C2 and\,D2 with BH companions (Fig. \ref{bin_evo_1_bh}) the orbital decay rate is 
about one order of magnitude lower, with $\dot{P}/P\approx -6\times 10^{-7}\mathrm{yr}^{-1}$. 
If the H/He gradient is shallower, as in Model\,B1 (Fig. \ref{ulxbh}), the orbital decay 
is faster by more than one order of magnitude ($\dot{P}/P\approx -10^{-5}\mathrm{yr}^{-1}$).

\cite{2015A&A...577A.130F} compiled and evaluated the orbital period changes of ten eclipsing SGXBs from a multi-decade monitoring campaign. 
Five sources in their sample showed a significant change of the orbital period. 
All of these five sources showed decaying orbits ($\dot{P}/P<0$), with rates $|\dot{P}/P| \sim 1\,...\,3\times 10^{-6}\,\mathrm{yr}^{-1}$. 
\begin{figure}[tb!]
\includegraphics[width=0.49\textwidth]{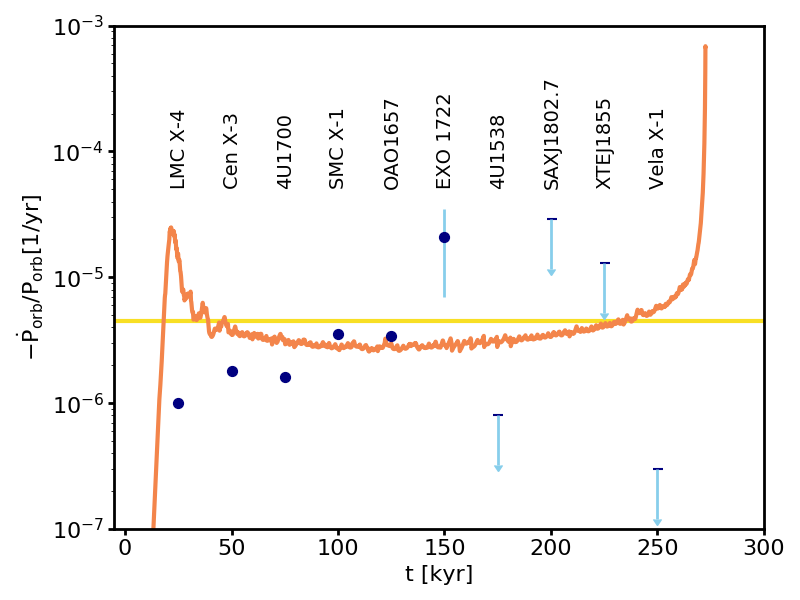}
\caption{Evolution of the orbital period derivative $\dot{P}/P$ of our Model\,D2 shown in Fig.\,\ref{D2} 
as given by the numerical simulation (orange line),  together with our analytic estimate for the same model (see text; yellow line).
Overplotted are the empirical decay rates of ten SGXBs as inferred by \cite{2015A&A...577A.130F},
who give six measured values with 1$\sigma$ error bars (blue dots; for most of these, the error bar is smaller than the size of the dot)
and four upper limits (blue arrows). We note that the time axis has no meaning for the period derivatives of the observed sources.}
\label{ppdot_fig}
\end{figure}
The source EXO 1722-363 shows an orbital decay rate of $\dot{P}/P=(-21\pm14)\times 10^{-6}\,\mathrm{yr}^{-1}$. 
Although the uncertainty is smaller than the measured value, \cite{2015A&A...577A.130F} did not label this source as significant. 
We note that the orbital decay rate is consistent with zero within the $2\sigma$ error.
For four more sources, no orbital decay within the $1\sigma$ environment was observed.
We used the $1\sigma$ uncertainty provided by \cite{2015A&A...577A.130F} as an upper limit for the orbital decay rate.

In Fig.\,\ref{ppdot_fig}, we compare the empirical values with the $\dot{P}/P$-evolution of our Model\,D2 with a NS accretor,
together with our analytic estimate for its period derivative. 
All five sources for which an orbital decay rate was measured with high significance are in good agreement with the decay rate 
of our model. I.e., deviations between the theoretical model and the observed orbital decay rate are 
not much larger than a factor of two (note that the time coordinate for the empirical values in Fig.\,\ref{ppdot_fig} has no meaning). 
This is remarkable, given that our models were in no way tailored to reproduce the observed sources.

The only source that does not fit well to our model is Vela X-1. 
The upper limit of the decay rate lies one order of magnitude below our prediction. Equation\,\ref{pdotpa} 
suggests that the mass-transfer rate must not exceed a few times $10^{-7}\mathrm{yr}^{-1}$ in order to obtain
a decay rate below the upper observation limit. We did not find such low mass-transfer rates in our models with NS accretors. 
We must hence conclude that the orbital change of Vela X-1 is at least not modelled with our simple prescription. 
This does not completely rule out Roche-lobe overflow as a possible accretion mode for this source. 
The small orbital decay rate may be explained, if Vela X-1 was just in a transition state between 
wind dominated accretion and atmospheric Roche-lobe overflow.

We note that orbital decay may be more complex as discussed in this section. In our simplified description, 
effects like tidal interaction \citep{1976ApJ...205..556L,1984A&A...135..155V,1996ApJ...456L..37S,2000ApJ...541..194L} 
or the Darwin instability \citep{1994ApJ...420..811L} are neglected. These mechanisms may also be able to drive orbital decay 
that is in agreement with the observed decay rates \citep{1983ApJ...268..790K,1984A&A...135..155V,
1993ApJ...410..328L,1996ApJ...459..259R,1996ApJ...456L..37S,2012ApJ...759..124J}.
In any case, our calculations show that the observed orbital decay rates of SGXBs may be explained by 
the simple isotropic re-emission model, as long as the mass transfer occurs on the nuclear timescale. 
The fact that highly non-conservative and conservative mass transfer show the same orbital decay rates 
for high mass ratios implies that SGXBs and ULXs should show the similar values of $\dot{P}/P$ within this
prescription.

\section{Binary evolution models for ULXs}
\label{Sbinary}

The Eddington luminosity of a $10\,\mathrm{M}_\sun$ black hole is about $3\times 10^{39}\egs$. Brighter X-ray sources are
considered as ultra-luminous X-ray sources (ULXs). As the nuclear timescale mass-transfer rates in our models
are in the range $10^{-5}\dots 10^{-6}\msoy$, some of them might be considered as models for ULXs with up to
$\sim 10^{41}\egs$ (cf., Tab.\,\ref{tab}), if all the transferred matter would be accreted.  

\subsection{ULXs with neutron star}

It has recently been discovered that several ULXs show X-ray pulsations, which is only expected
if the compact accretor is a neutron star \citep{2017ARA&A..55..303K}. Whereas the Eddington limit of neutron stars
is well below $10^{39}\egs$, some of the X-ray pulsating ULXs show X-ray luminosities which are clearly in the
ULX regime \citep{2014Natur.514..202B}. 
Whereas beaming effects might help to explain the very large X-ray luminosities \citep{2001tysc.confE1012K}, some neutron stars in ULXs 
have been shown to experience an extreme spin-up on a short timescale, which may require actual accretion
onto the neutron star at rates  much above the classical Eddington-limit (\citeauthor{2017Sci...355..817I} \citeyear{2017Sci...355..817I}, but see \citeauthor{2019arXiv190303624K} \citeyear{2019arXiv190303624K}).
One way to understand such high luminosities from accreting neutron stars is to invoke 
magnetic fields with a field strength above $\sim 10^{13}\,$G, which could reduce the radiative opacity
of the accreted matter and thus raise the Eddington limit \citep{2017Sci...355..817I}.


ULXs hosting neutron stars have also been suggested to host intermediate mass donor stars.
However, \cite{2017ApJ...846..170T} argued that corresponding models only work for donor star masses below $7\mso$.
The donor star in the ultra-luminous X-ray pulsar in NGC\,5907\,ULX is constrained to be larger than $\sim 10\,\mathrm{M}_\sun$,
and its orbital period is found to be $P_{\rm orb}=5.3^{+2.0}_{-0.9}\,$d \citep{2017Sci...355..817I}.
\cite{2014Natur.514..198M} determined the orbital period of the ULX pulsator NGC\,7793\,P13
to $\sim 64\,d$, where the mass donor is a B9Ia supergiant of about $20\,\mathrm{M}_\sun$ \citep{2017MNRAS.466L..48I}.
Except for being ultraluminous, their parameters are reminiscent of those of the SGXBs. 

\begin{figure}[tb!]
\includegraphics[width=0.49\textwidth]{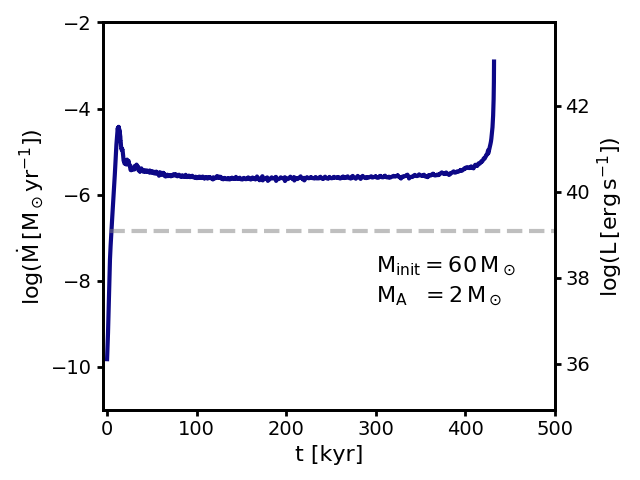}
\caption{Evolution of the mass-transfer rate for the same model as shown in Fig.\,\ref{D2} (left Y-axis) (Model\,D2),
but here calculated assuming conservative mass transfer. }
\label{ulx}
\end{figure}

We have seen from our Model\,D2 in Sect.\,\ref{modeld2} it obtains a nuclear timescale mass-transfer 
phase, with accretion rates which would lead to X-ray luminosities above $10^{40}\egs$ if the neutron star
could accrete all the matter. In order to probe this situation, we have repeated the calculation
displayed in Fig.\,\ref{D2}, but allowed the neutron star to accrete all of the transferred matter.
While this may not be realistic, as a fraction of the transferred matter may always be expelled,
it provides the limiting case, with more realistic models bounded by this and the Eddington limited 
calculations shown in Sect.\,\ref{modeld2}.

Figure\,\ref{ulx} shows that the long duration of the X-ray bright mass-transfer phase
is not only maintained by the conservative accretion model, but the time span of X-ray emission
is almost doubled, compared to the Eddington limited model. This is understandable since
the mass of the neutron star is growing significantly here -- it would likely collapse into
a black hole eventually, as a total of about $1.4\mso$ are transferred -- and the thereby reduced mass ratio leads to slightly smaller mass
transfer rates. However, its X-ray luminosity could still be well over $10^{40}\egs$ for more than
400\,000\,yr.

Our models do not attempt to reproduce any of the observed ULXs. However, they 
show that ULXs with neutron star accretors may may accrete from Roche-lobe overflow for 
much longer than a thermal timescale of the donor star, 
if their donor star is a helium-enriched supergiant.

\subsection{Black hole companions}

Clearly, ULXs can form in binaries when one component is a sufficiently massive black hole and the 
companion transfers mass at a high enough rate. In this situation, it has been recognised
that either beaming \citep{2001tysc.confE1012K, 2002A&A...382L..13K}, photon-bubbles \citep{2002ApJ...568L..97B, 2003ApJ...586..384R}, or magnetic accretion disc coronae \citep{2006ApJ...651.1049S}
could help to raise the apparent or true Eddington limit, such that ULXs of up to
$10^{41}\egs$ can be explained with stellar mass black holes \citep{2008ApJ...688.1235M,2017A&A...604A..55M}.

\begin{figure}[tb!]
\includegraphics[width=0.49\textwidth]{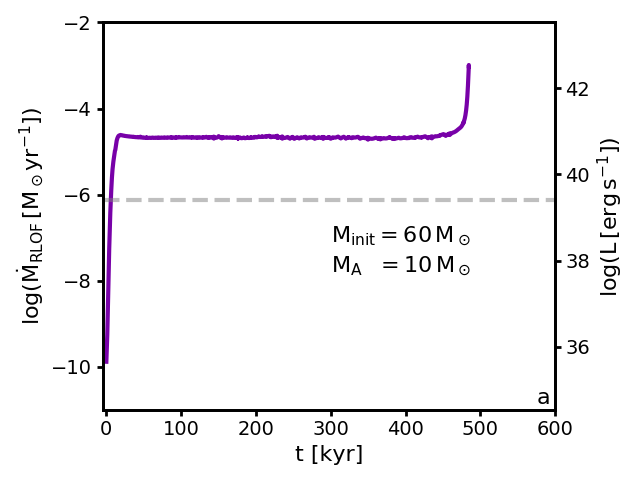}
\caption{Evolution of the mass-transfer rate as function of time for our Model\,B1,
which starts with a $42.5\,\mathrm{M}_\sun$ supergiant and a $10\,\mathrm{M}_\sun$ BH in an 18\,d orbit.}
\label{ulxbh}
\end{figure}

In order to obtain non-negligible ULX lifetimes, in these models the donor star is usually 
of comparable mass or smaller than the black hole, which severely limits the expected number of ULXs. 
In our models with helium-enriched donor stars, this restriction can be dropped. Assuming a $10\,\mathrm{M}_\sun$
black hole, the initial mass ratio expected in such systems is significantly smaller than
in the case of a neutron star companion. Consequently, we find nuclear timescale mass transfer
in more system if a black hole is assumed to be present.
In Fig.\,\ref{ulxbh}, we highlight our Models\,B1 with a donor star of initially $42.5\,\mathrm{M}_\sun$,
which provides a high mass-transfer rate to an initially $10\mso$ black hole for almost 0.5\,Myr, and which could provide
an X-ray luminosity of $\sim 10^{41}\,\mathrm{erg/s}$ if super-Eddington accretion is assumed.

In our models with neutron star accretors, only donor stars with steep H/He-gradients led to
nuclear timescale Roche-lobe overflow phases. As the five times smaller mass ratio
in systems with $10\,\mathrm{M}_\sun$ black hole accretors leads to a slower shrinking of the orbit,
even the ten times shallower H/He-gradient is sufficient to achieve nuclear timescale Roche-lobe overflow
in these systems. This may open the parameter space for BH-ULX systems significantly.

\section{Implications for origin of SGXBs}

The way the donor stars of our supergiant and ultra-luminous X-ray binary models have been fabricated
may raise doubts about their applicability to interpret the observed systems. In particular, we have seen that
nuclear timescale mass transfer was only achieved when the chemically homogeneous part of the hydrogen-rich envelop 
was removed {\em before} the mass transfer to the compact companion starts. This raises the question whether or not this can
occur in reality. 

\subsection{Clues from observations}



\begin{figure*}[htbp!]
\includegraphics[width=0.9\textwidth]{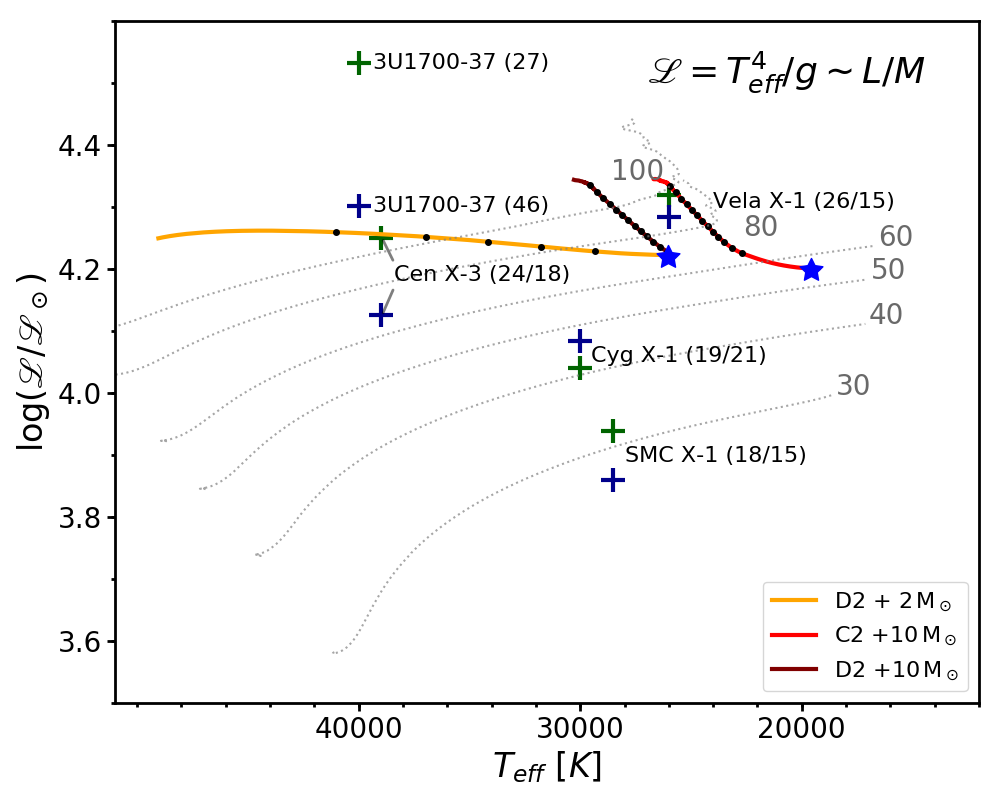}
\label{spec_hrd2} 
\caption{
Spectroscopic HR diagram including data for SGXBs from \cite{1978A&A....63..225C} (green crosses) and \cite{2015A&A...577A.130F} (blue crosses). 
The name of the source is labelled to the symbols. The parentheses include the donor masses as measured by (\cite{1978A&A....63..225C} / \cite{2015A&A...577A.130F}).
The gray dashed lines are tracks from single star evolutionary models \citep{2011A&A...530A.115B}. The gray numbers indicate the initial mass of the corresponding stellar model.
Red and orange lines indicate the evolution of the discussed binary models. The starting point of each models is labelled by a blue star. The black dots correspond to time steps of 50 000 years.}
\end{figure*}

The first idea, i.e., that the H/He-transition layer is close to the surface of the donor star, appears to be supported
by several observations. Using the data of \cite{1978A&A....63..225C} and \cite{2015A&A...577A.130F}, 
who determined the effective temperatures and surface gravities of several SGXB donors, we can compare the with stellar
models in a spectroscopic HR diagram (Langer \& Kudritzki, 2014). In this diagram, the ordinate values are proportional to
the luminosity-to-mass ratio of the stars. As already noted by \cite{1978A&A....63..225C}, Fig.\,\ref{spec_hrd2} shows that the SGXBs
do not match single star tracks of the corresponding mass. For instance, the donor star in Vela\,X-1 appears close to the
$100\,\mathrm{M}_\sun$ track, while its mass, inferred from radial velocity measurements, is only $\sim 25\,\text{M}_\odot$. 

As removing the stellar envelope does decrease the mass, but does not affect the luminosity significantly, it is not surprising 
that some of our SGXB models fit the sHRD position of Vela\,X-1, and other SGXB donors, better than the single star models.
I.e., Fig.\,\ref{spec_hrd2} shows that Model\,C2+$10\,\mathrm{M}_\sun$ gets close to the position of Vela\,X-1 near the end of its
mass-transfer phase, where the mass of the donor star is about $33\,\mathrm{M}_\sun$ (cf., Tab.\,\ref{tab}), suggesting an initial donor mass
closer to $60\,\mathrm{M}_\sun$. Whereas Vela\,X-1 has no black hole but a neutron star companion, this comparison shows that the
$L/M$-ratio of our model may be close to that of the donor star in Vela\,X-1, implying that it did lose its hydrogen-rich envelope
at an earlier stage of its evolution. Figure\,\ref{spec_hrd2} shows that in all cases for which effective temperature and
surface gravity could be determined, the corresponding $L/M$-ratio is well above that of the single star models. 
This indicates that all the donor stars have already lost a significant fraction, and perhaps all, of their non-enriched 
massive envelope. 

An even more direct evidence for this comes from
model atmosphere calculations and fits to observed donor star optical spectra 
of the SGXBs 4U 1700-377 \citep{2002A&A...392..909C}, GX301-2 
\citep{2006A&A...457..595K} and Vela X-1 \citep{2018A&A...610A..60S}, who find an enhanced surface helium abundance in all three cases.
A surface helium enrichment is expected to occur only after the chemically homogeneous, non-enriched part of the stellar envelope is removed.
The implication of the observed helium enrichments is that due to mass loss, whether by stellar winds or by Roche-lobe overflow,
the donor star radii are currently decreasing. The remarkable circumstance that, nevertheless, the donor star in these systems
are very near Roche-lobe filling may imply that the orbit shrinks at the same time. This is only expected if the systems were 
currently undergoing Roche-lobe overflow.   

We conclude that SGXB observations provide ample of evidence in support of the assumption that
their donor stars have lost the non-enriched massive hydrogen envelope in a previous phase of evolution.



\subsection{Clues from stellar models}

The question to ask at this stage is through which mechanism the donor stars in SGXBs lost their H-rich envelopes
before they entered the X-ray binary stage. As their initial masses appear all very high, one may wonder whether
the ordinary radiation driven winds of massive stars are sufficient to reach this goal. 
Looking at the massive star models in the literature \citep{2014ARA&A..52..487S, 2011A&A...530A.115B, 2001A&A...369..574V}  and considering that the wind mass loss rates
for these very massive stars can not be predicted better than within a factor of$\,\sim 2$ (with e.g., some of this coming from the metallicity
spread in the Galaxy), it may not be a problem to remove the required amounts of mass by stellar winds.

However, this mechanism would require a significant amount of fine tuning.  
Since stellar wind mass loss always widens the orbits of binary stars (cf., Eq.\,\ref{orbita} in Sect.\,\ref{srr}),
the OB\,star's expansion needs to catch up with the increasing Roche-lobe radius
just at the time where H/He-gradient appears near the stellar surface. Figure\,\ref{response_rl} shows that
this is not impossible, since the massive star models tend to increase in size as mass is being removed
until shortly before helium-enriched layers appear at the stellar surface (see also Fig.\,\ref{cee_d40m10w0} below).
But only systems within a narrow initial period range would be able to fulfil the timing constraint.

It may be the more common situation that the donor star fills its Roche radius
at a time when the H/He-interface layer is still buried beneath a massive hydrogen-rich envelope. As we have seen in Sect.\,\ref{Sbinary},
this leads to very high mass-transfer rates and likely to a common envelope evolution (see also \cite{1987ApJ...318..794H}). 
Here, the Roche-lobe overflow could start in the advanced phase of core hydrogen burning, or after core hydrogen exhaustion.
While it is beyond the scope of this paper to comprehensively investigate the outcome of such an evolution, we provide a simple estimate
as follows.

\begin{figure}[htbp!]
\includegraphics[width=88mm]{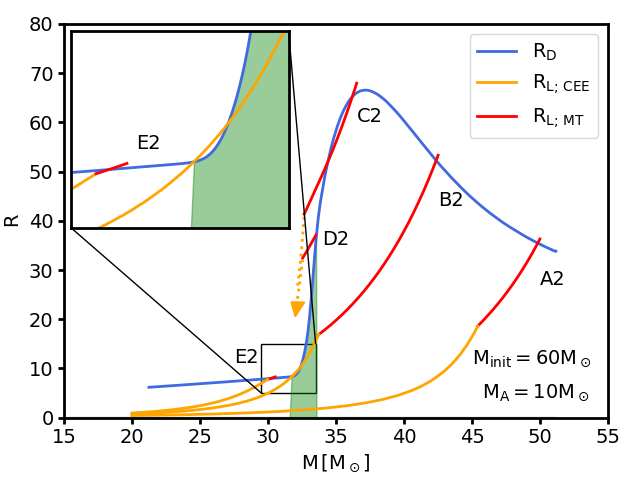}
\caption{Donor radius as function of mass in thermal equilibrium (blue). Red lines show the Roche radius as function of donor mass during the binary calculation including a $10\,\text{M}_\sun$ accretor and no stellar wind of the donor. The yellow lines show the Roche radius after the binary calculation has stopped and CEE is initiated. The orbital separation and hence the Roche radius was inferred using the energy budget description of CEE. The green area marks the position of the H/He-transition layer.}
\label{cee_d40m10w0}
\end{figure}

We show the radius evolution of our $60\,\mathrm{M}_\sun$ model (blue line), assuming nuclear timescale mass loss, in Fig.\,\ref{cee_d40m10w0}, together with
the evolution of the Roche radius for our Sequences\,A2$\dots$E2 during the mass-transfer evolution (red lines). 
At the end of the mass-transfer evolution, a common envelope phase is expected. The yellow lines in Fig.\,\ref{cee_d40m10w0} show the value of the
donor's Roche-lobe radius at a given donor star mass if the common envelope would be removed at the corresponding time, where the 
Roche-lobe radius is obtained from
equating the energy release $\Delta E$ from the decaying black hole orbit with the envelope binding energy $E_\text{bin}$ of the envelope above this orbit where 
\begin{equation}
\Delta E=-\frac{G M_\text{D, i} M_\text{A}}{2a_\text{i}} +\frac{G M_\text{D} M_\text{A}}{2a} 
\label{cee1}
\end{equation}
and
\begin{equation}
E_\text{bin}=\int_ {M_\text{D}}^{M_\text{D, i}}-\frac{Gm}{r}+u\,\mathrm{d}m
\label{cee2}
\end{equation}
is the effective binding energy, which includes gravitational binding energy reduced by the thermal energy. 
Here, $M_\text{D, i}$ is the donor star's mass at the beginning of the common envelope evolution, $M_\text{D}$  its mass at a putative end stage
of the common envelope evolution, and $a_\text{i}$ and $a$ the corresponding orbital separations.
The Roche radius during RLO was directly inferred from the binary calculation. At the point where the binary calculation stops, we use the last calculated donor model,
orbital separation and accretor mass to compute the Roche radius as function of the donor mass as described above.
The stellar model of the donor at the beginning of the common envelope evolution defines the envelope binding energy $E_\text{bin}$ as function of 
the remaining mass $M_\text{D}$. The condition of Roche-lobe filling at the beginning of the common envelope evolution sets the orbital separation at that time
for a given accretor mass. Assuming the accretor mass remains constant allows us to compute the Roche radius after the common envelope evolution 
as a function of donor mass $M_\text{D}$ at that time. 

The evolution of our Sequence\,A2 in Fig.\,\ref{cee_d40m10w0} provides a case where a merger appears to be the most likely outcome
of the common envelope evolution. At its onset, the donor star's mass is about $45\,\mathrm{M}_\sun$ and its H/He-transition layer is buried beneath more than $10\,\mathrm{M}_\sun$ 
of hydrogen-rich envelope. At the same time, it is rather compact ($R_{\rm D}\simeq 20\rso$) due to its high mass-transfer rate. 
The yellow line for Sequence\,A2 in Fig.\,\ref{cee_d40m10w0} shows that there is no possible final mass after the common envelope evolution for which
the donor's Roche-lobe radius would exceed its thermal equilibrium radius. While the donor's radius may be smaller than its thermal equilibrium radius
during mass transfer or immediately after a common envelope ejection, the implication is that if at all, it would be able to fit into its Roche radius
only for a thermal timescale or less. Afterwards, it would expand and merge with the companion.

However, we see a different picture for Sequence\,B2, which is also expected to quickly undergo a common envelope evolution (cf., Fig.\,\ref{bin_evo_1_bh}).
Figure\,\ref{cee_d40m10w0} shows that in this case, the yellow line indicating the donor's Roche-lobe radius crosses the blue line for the donor's thermal equilibrium
radius. Consequently, this model opens the possibility of a successful common envelope ejection at a time
where the donor star's H/He-transition layer is at the stellar surface. After this, the expectation is that the nuclear timescale expansion of the
donor would make it fill its Roche-lobe soon again, allowing nuclear timescale mass transfer onto the compact companion.

While our estimate in Fig.\,\ref{cee_d40m10w0} includes many simplifications and is not to be understood as a quantitative model,
it shows the interesting possibility to interpret SGXBs as post-common envelope systems. In this frame, the emerging of the
H/He-transition layer at the donor's surface at the end of the common envelope evolution is not a matter of fine tuning,
but is naturally produced due to the sharp drop of its thermal equilibrium radius at this time. Potentially, all the models with
initial masses in between those of Sequences\,B2 and\,C2 could follow this path. 
The evolution of Sequence\,B2 shows that in this scenario, a significant fraction, or even the major fraction, of the hydrogen-rich envelope
may be transferred to the compact companion in a thermal timescale mass-transfer event before the onset of the common envelope
evolution. The higher this fraction is, the higher are the chances to avoid a merger during the common envelope evolution.

We may point out that there are tentative observational counterparts for this type of evolution. I.e., the enigmatic X-ray binary SS\,433
appears to have a supergiant mass donor providing mass at the thermal mass transfer rate \citep{2004ASPRv..12....1F}. Within our picture,
SS\,433 may qualitatively correspond to our Models\,B2 or \,C2 at the time of the first mass-transfer peak (cf., Fig.\,\ref{bin_evo_1_bh}). It would then
evolve into an ordinary SGXB after going through a common envelope phase (Model\,B2), or avoiding a common envelope phase (Model\,C2).

Finally, our common envelope scenario for the pre-SGXB evolution may relate to the so called obscured SGXBs which have been discovered
recently \citep{2013AdSpR..52.2132C}, as the ejected envelope may provide enough circumstellar material to produce the obscuration.


\subsection{Metallicity dependence}

\begin{figure}[htbp!]
\includegraphics[width=0.37\textwidth,angle=270]{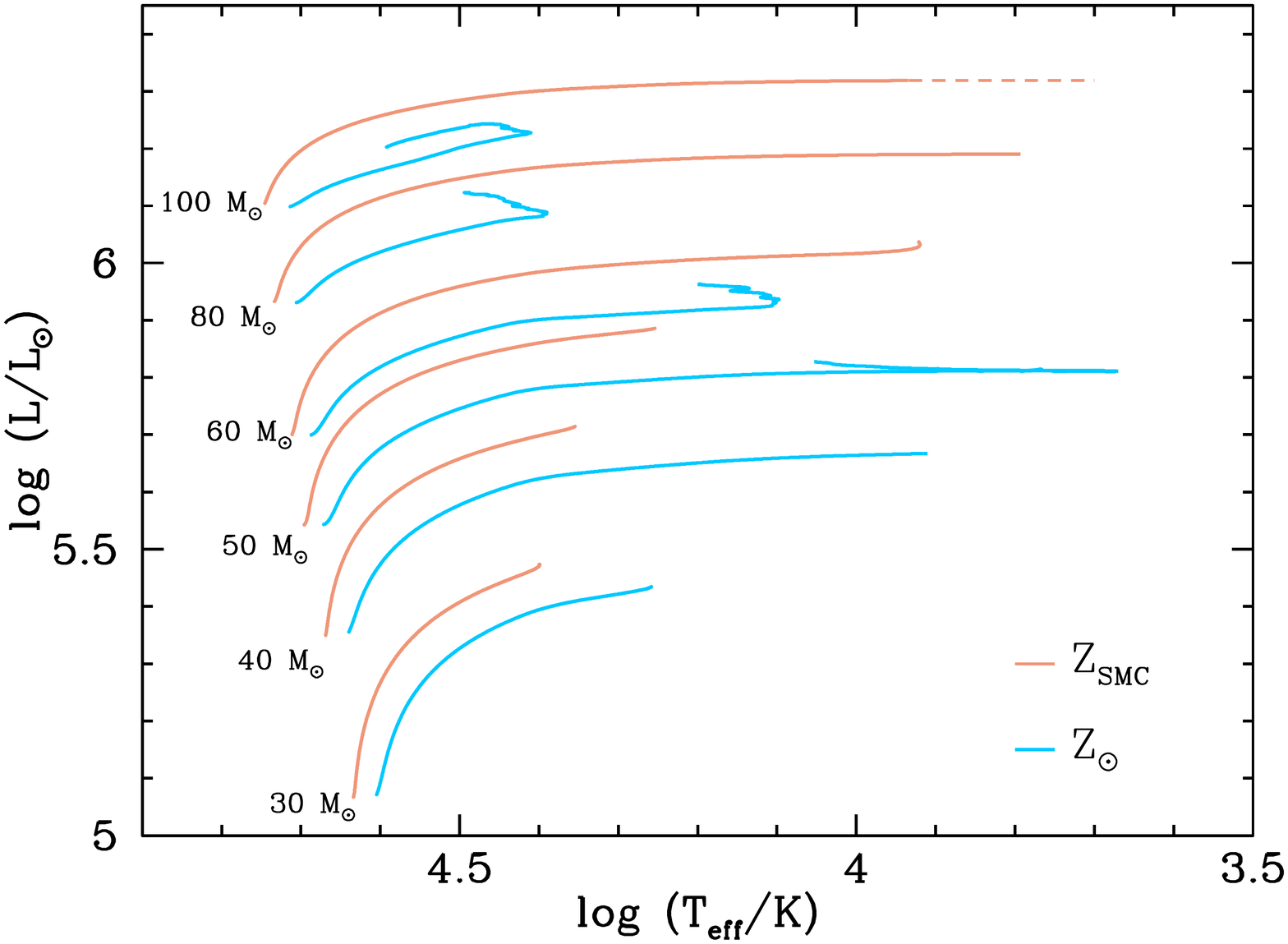}
\caption{Evolutionary tracks for Solar and SMC metallicity during core hydrogen burning \citep{2011A&A...530A.115B}.}
\label{hrd2z}
\end{figure}

We want to point out that the scenario for producing the SGXBs, either due to well-timed stellar wind mass loss, such that it starts
Roche-lobe overflow when the H/He-transition layers appear at the surface, or through a pre-mass-transfer common envelope phase,
may be favoured in a high metallicity environment like our Galaxy. This becomes clear when we compare the evolutionary tracks of
massive single stars for Solar metallicity with those computed with an initial composition as that in the Small Magellanic Cloud
(Fig.\,\ref{hrd2z}). It shows that the first scenario does not appear to be available at low metallicity. While at solar metallicity,
the models lose enough mass to expose their H/He-transition layers without the help of a companion star, the low-metallicity models
never do that. 

In the Milky Way, it has been found by \cite{1979ApJ...232..409H} that the most massive stars avoid the upper right part of
the HR diagram, which has been related to the Eddington limit and the instabilities in the so called Luminous Blue Variables 
\citep{1988ApJ...324..279L, 1998ApJ...504..200U, 2012A&A...538A..40G, 2015A&A...580A..20S}. The proximity of the Galactic supergiants to the
Eddington limit will therefore likely facilitate the loss of the hydrogen-rich envelope also when a compact companion is present.
Due to the iron opacity, the phenomenon is shifted to much higher luminosities and masses for low metallicities 
\citep{1998ApJ...504..200U, 2017A&A...597A..71S}. Therefore, the mass of the hydrogen-rich envelop in SMC supergiants
with compact companions, at the time they would likely fill their Roche-lobe, is likely much larger than in a comparable case in the
Galaxy, and an immediate merger may be expected. 
While in the Milky Way, we have about as many SGXBs as Be/X-ray binaries \citep{2015A&ARv..23....2W}, only one SGXB is known in the SMC, but 81 confirmed Be/X-ray binaries \citep{2016A&A...586A..81H}.
Our model may offer a natural explanation of this strong metallicity dependence.

A metallicity dependence would also be expected for the ULX-application of our scenario. In fact, what has be said about the 
donor stars in SGXBs holds the same for ULX supergiant donors. This is particular relevant for ULXs with neutron star accretors,
as in those the mass ratio problem is the strongest. Consequently, we would expect ULXs with neutron stars preferentially at high
metallicity. Our common envelope scenario may also have implications for understanding the generation of large scale magnetic fields
in stars. I.e., based on the observation that magnetic white dwarfs are common in Cataclysmic Variables, but absent in wide binaries
\citep{2000PASP..112..873W}, \cite{2010MNRAS.402.1072P} argue that the strong magnetic field in CVs with
a magnetic white dwarf could be generated during their previous common envelope evolution. While their model appears only marginally
successful, the idea may still apply. If it was transposable to massive X-ray binaries, it could imply that a common envelope phase 
in a pre-SGXB evolution may help to induce a strong field into the neutron star, with the effect of producing a ULX rather than an
ordinary SGXB.

\section{The fates of SGXBs and SG-ULXs}
\label{fates}
We have ended our binary evolution calculations when either the mass-transfer rate exceeded $\sim 10^{-2}\msoy$, or when
the models became thermally unstable due to core hydrogen exhaustion. Therefore, we can only conjecture about the
further evolution of our model binaries, and about the fates of the observed SGXBs and SG-ULXs.

From the time dependence of the mass-transfer rates of our models, it can be concluded that
after the H/He-transition layer is lost during the nuclear timescale evolution, the mass-transfer rate increases 
steeply. This is the case for our Models\,B1 and\,C1 with black hole accretors (Fig.\,\ref{bin_evo_1_bh}),
and for Model\,D2 with a neutron star accretor (Fig.\,\ref{bin_evo_1_ns}), also in the case of conservative
mass transfer (Fig.\,\ref{ulx}). According to \cite{1987ApJ...318..794H}, who inferred a maximum mass ratio of $2.14$ for a binary to be stable against dynamical runaway, the onset of a common envelope evolution in these models 
appears likely. Since at this stage, our models are rather compact and still possess a significant envelope with a hydrogen mass fraction of 20\%,
we assume that a merger would be the likely outcome. 

In another suite of our models, the nuclear timescale mass transfer lasts until hydrogen is exhausted in the core of the
donor. The ensuing contraction of the donor model leads to a sharp drop of the mass-transfer rate, and the Roche-lobe overflow
phase stops (Model\,D2 with neutron star accretor and constant wind mass loss, Fig.\,\ref{bin_evo_1_ns}; and Models\,C2
and\,D2 with black hole accretors, Fig.\,\ref{bin_evo_1_bh}). We would expect that a hydrogen shell source will ignite in
these models, which will expand their envelops and initiate a thermal timescale mass transfer. Again, using
the criterion of \cite{1987ApJ...318..794H}, we would expect this to evolve into a common envelope phase, with a merger
of both components as the result. 
Only two of our models with a time dependant stellar winds (Models\,B2 and\,C2 with black hole accretors) are expected to
emerge as a Wolf-Rayet$-$black hole binary.

However, our models are fabricated and are not self-consistently evolved from the zero-age main sequence stage.
For Models\,A1 to\,E1, whose transition layer contains the shallower 
H/He-gradient, the slope is naturally expected from the receding convective core during hydrogen burning.
However the steeper H/He-gradient in Models\,A2 to\,E2 is perhaps only formed after core hydrogen exhaustion (Schootemeijer \& Langer 2018).
In core helium burning supergiants, the nuclear timescale mass transfer might then stop at the time of core helium exhaustion.
It may not be excluded that in this case, a merger during the short remaining time to the collapse of the donor star
can be avoided, and a short period double compact binary emerges. 

Clearly, a further investigation of the post-mass-transfer
evolution of more self-consistent SGXBs and SG-ULXs appears warranted.

\section{Conclusions}

We refuted the long-standing paradigm that mass transfer in high mass ratio binary systems
must be unstable and can last for at most a thermal timescale of the donor star. We have first produced single star
models with mass radius exponents (e.g., Fig.\,\ref{m_response_he1}) well exceeding the ones required 
for stable mass transfer in high mass ratio binaries (Fig.\,\ref{response_rl}). We have then 
calculated detailed binary evolution models
for SGXBs with neutron star and black hole components, and found nuclear timescale mass transfer in many of them,
with initial donor to accretor mass ratios of up to $\sim 20$.

The key for the stability of the mass transfer in our models despite the inevitably strong orbital contraction
is that their surface helium abundance is
increasing. Thereby, the donor stars deflate their radius by a factor of a few during the mass transfer, evolving from
supergiants into Wolf-Rayet stars. 

While our models are fabricated and not derived from earlier phases of binary evolution,
their key feature appears to be supported by observations, as the SGXB donors generally appear to be
helium-rich and overluminous for their mass (Fig.\,\ref{spec_hrd2}).   

Furthermore, our models make it easier to understand several observations. 
The first is the large number of observed SGXBs in the Milky Way, which is proportional to the
duration of their X-ray phase. The latter is drastically extended by our models.
The second is the discovery of ULXs with supergiant donor stars and neutron star accretors. 
Again, in the standard picture, their life time is expected to be extremely small. We show an example
with an initial mass ratio of 17 and a ULX phase due to stable Roche-lobe overflow lasting
for more than 400\,000\,yr. 

We argue that the SGXBs in the Milky Way, and perhaps also the ULXs with neutron star accretors may have formed
in a common envelope evolution, during which the loss of the H-rich envelope of the donor star 
was eased as it reached its Eddington limit. Such a common envelope phase might be related
to the dense circumstellar medium found in the Galactic obscured SGXBs.
While this could remove the fine-tuning problem to produce the SGXBs in their current state, 
it could also explain the nearly complete lack of SGXBs in the SMC, where stars reach their
Eddington limit only at much higher mass. 

If ULXs with neutron star accretors would correspond to our common envelope scenario, they
would be preferentially expected at high metallicity, and be rare at low metallicity, analogous to
the SGXBs. ULXs with massive black hole accretors, on the other hand, might have a preference
for low metallicities, where the final black hole masses are expected to be larger.

This scenario may also shed light on the question how magnetar fields are created. It would make 
the ULXs with neutron star accretors, in which accretion rates of more than two orders of magnitude 
above the Eddington limit are thought to be enabled by extreme neutron star magnetic fields, 
the more massive cousins of polars, i.e., CVs in which a
main sequence star sheds mass onto a magnetic white dwarf, and which are also thought to have
undergone a previous common envelope evolution. 

It is necessary to produce detailed progenitor models for SGXBs in order back-up the common envelope scenario.
This may be lucrative as an understanding of the common envelope evolution of very massive stars
as function of metallicity is also required to reliably predict the population of double black hole binaries ---
into which some of the Galactic SGXBs might evolve. At the same time, may be several avenues to
provide stronger observational constraints to the common envelope evolution of very massive stars, including
the search for OB\,star$+$BH/NS binaries, of which many may have escaped detection so far.

\begin{acknowledgements}
We are grateful to Ed van den Heuvel and Philipp Podsiadlowski for useful comments and discussions. TMT acknowledges an AIAS-COFUND Senior Fellowship funded by the European Union's Horizon 2020 Research and Innovation Programme (grant agreement no 754513) and Aarhus University Research Foundation.
\end{acknowledgements}

\bibliographystyle{aa}
\bibliography{thebib}

\clearpage

\section*{Appendix A --- Mass-radius exponents for all stellar models}
\renewcommand{\thefigure}{A\arabic{figure}}
\setcounter{figure}{0}

Since the response of the stellar radius to slow mass loss is the key to obtain nuclear timescale
mass transfer in our supergiant X-ray binaries, we show here the corresponding mass-radius exponents $\zeta_{\rm R}$
for all of our stellar models. 

\begin{figure}[htbp!]
\includegraphics[width=88mm]{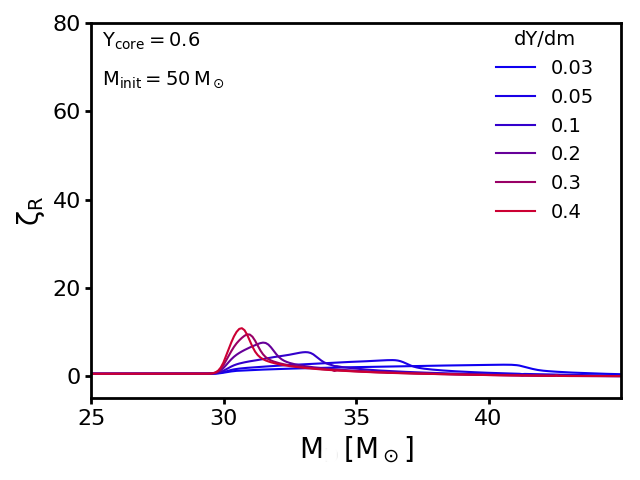}
\includegraphics[width=88mm]{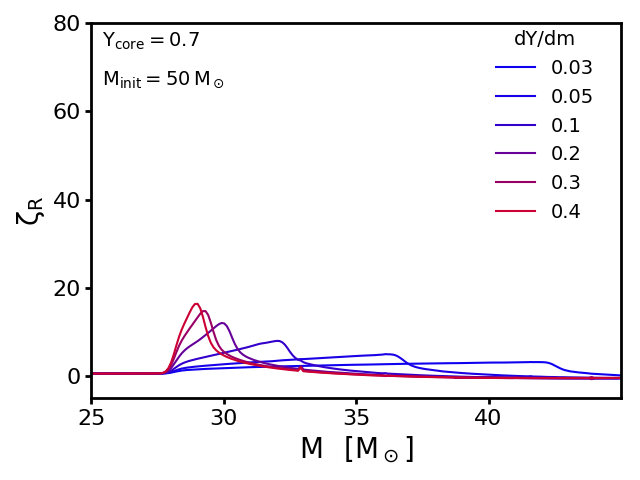}
\includegraphics[width=88mm]{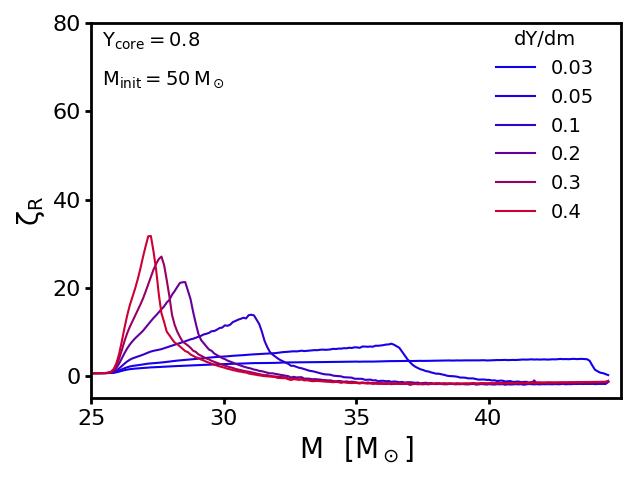}
\caption{Mass-radius exponent as function of mass for our initially \protect $ 50\ \mathrm{M_\sun}$ model,  
for different core helium abundances. }
\label{respall1}
\end{figure}

\begin{figure}[htbp!]
\vspace{1.4cm}
\includegraphics[width=88mm]{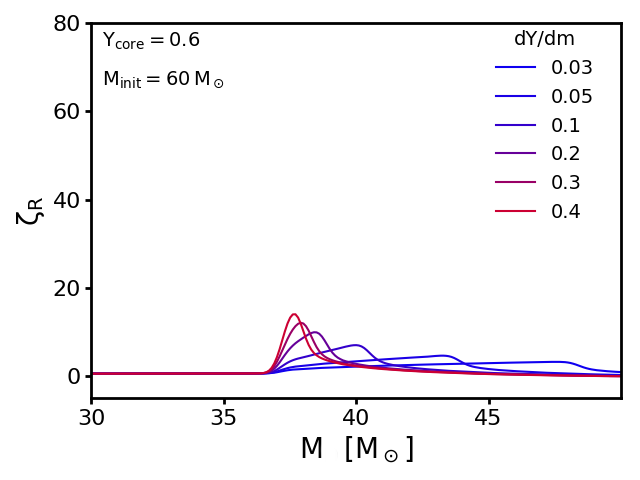}
\includegraphics[width=88mm]{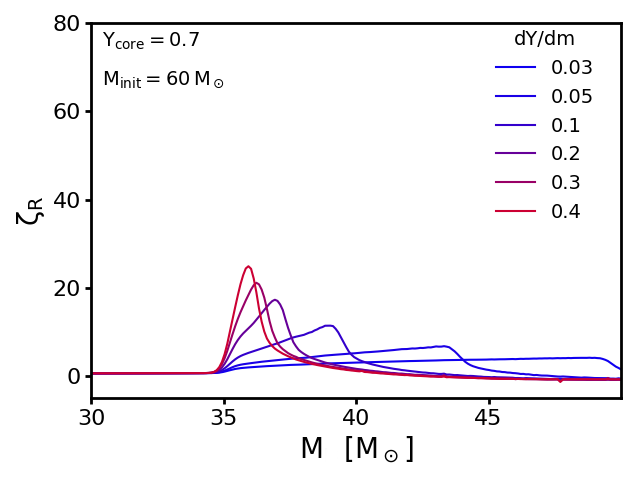}
\includegraphics[width=88mm]{fA2c.png}
\caption{\label{respall2} As Fig.\,\ref{respall1}, but for our 60$\mso$ model. }
\end{figure}
\begin{figure}[htbp!]
\includegraphics[width=88mm]{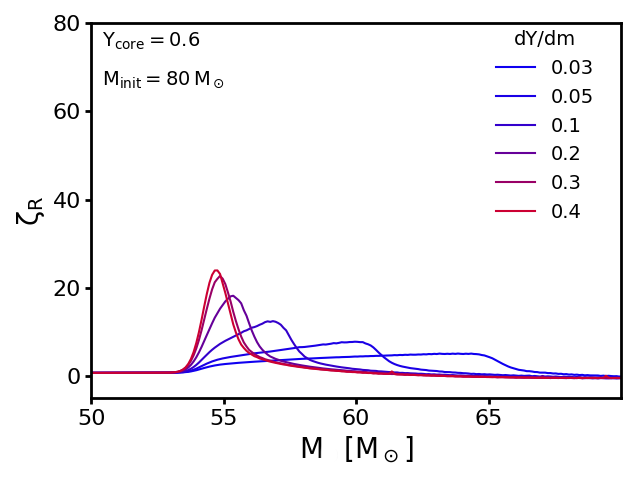}
\includegraphics[width=88mm]{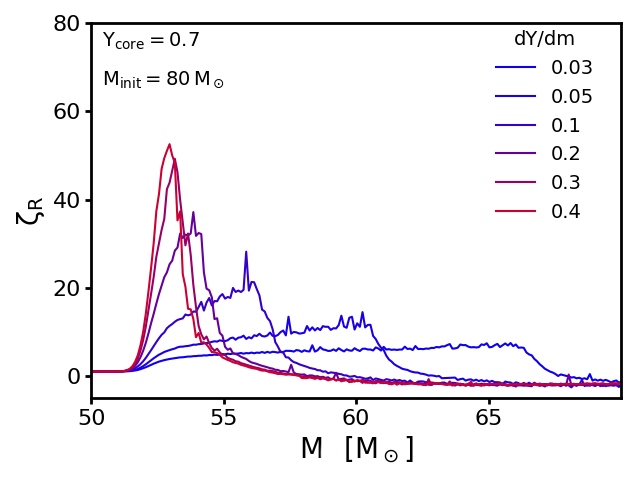}
\includegraphics[width=88mm]{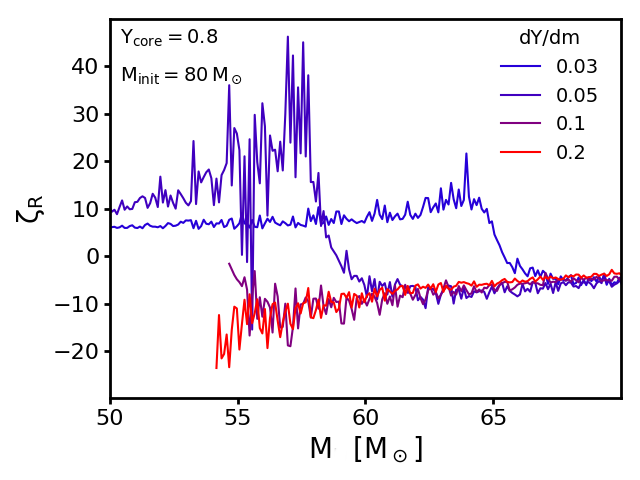}
\caption{\label{respall3} As Fig.\,\ref{respall1}, but for our 80$\mso$ model. 
For a core helium mass fraction of $Y_{\rm c}=0.8$ and steep helium profiles, we encountered convergence problems.}
\end{figure}

\clearpage

\section*{Appendix B --- Details for our binary evolution models}
\renewcommand{\thefigure}{B\arabic{figure}}
\setcounter{figure}{0}

\begin{figure*}[]
\includegraphics[width=0.49\textwidth]{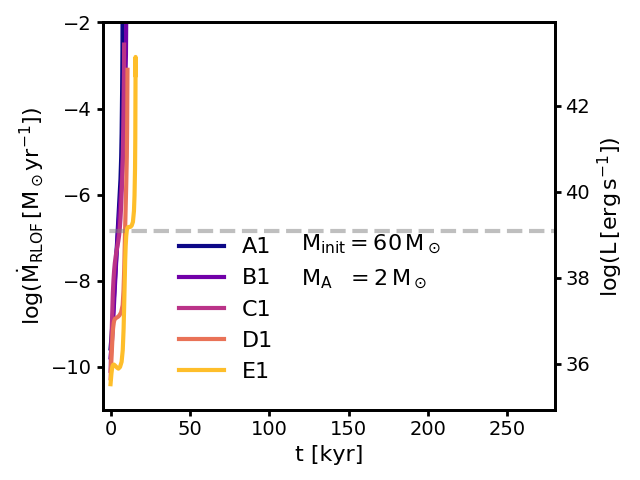}
\includegraphics[width=0.49\textwidth]{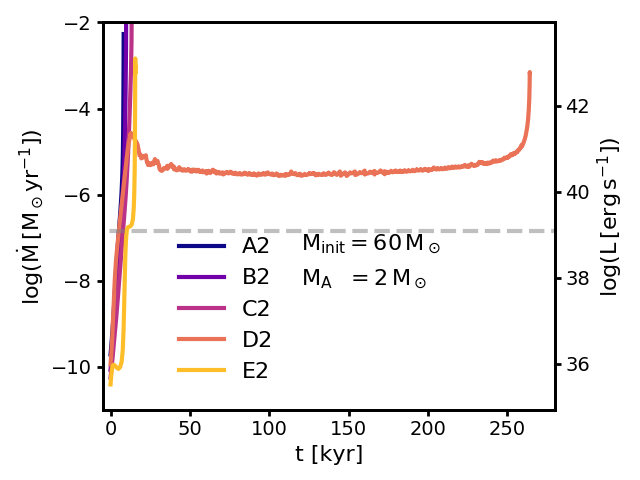}\\
\includegraphics[width=0.49\textwidth]{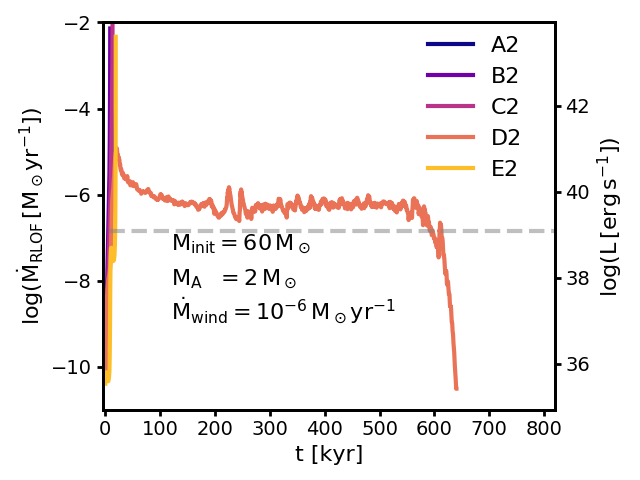}
\includegraphics[width=0.49\textwidth]{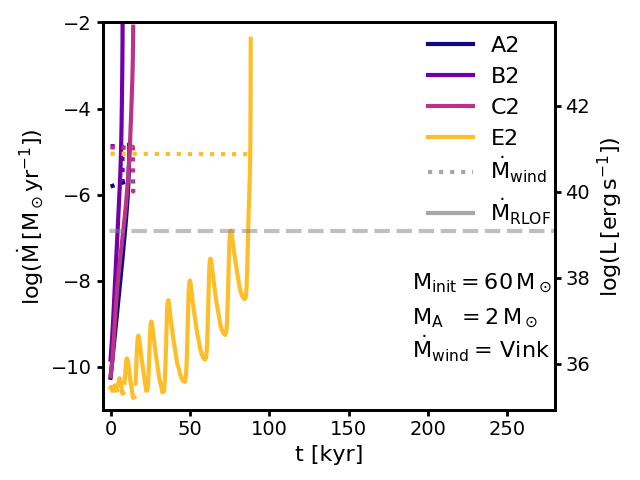}
\caption{\label{bin_evo_1_ns} 
Evolution of mass-transfer rate for three initial models and a NS accretor. The right axis in indicates the X-ray luminosity corresponding to the mass transfer rate (not Eddington limited) assuming an accretion efficiency of $\eta=0.15$. The dashed line indicates the Eddington accretion limit of the  accretor.}
\end{figure*}

\begin{figure*}[]
\includegraphics[width=0.49\textwidth]{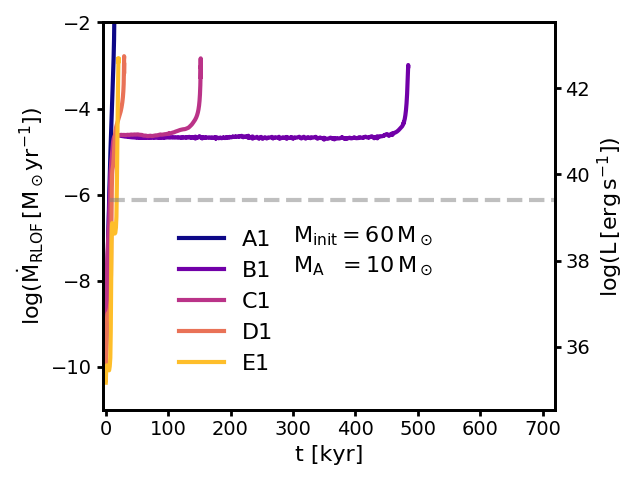}
\includegraphics[width=0.49\textwidth]{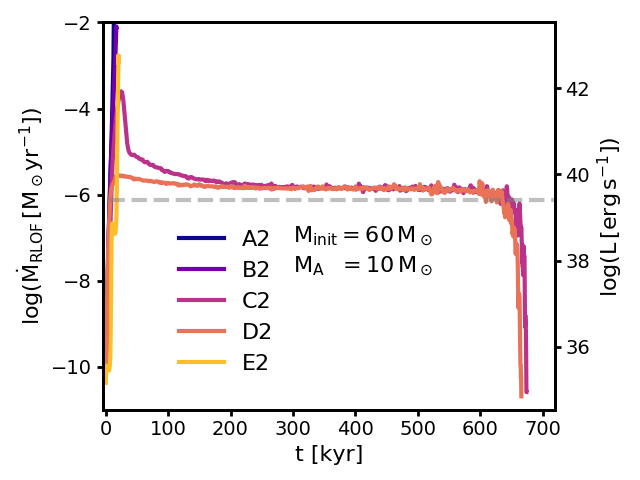}\\
\includegraphics[width=0.49\textwidth]{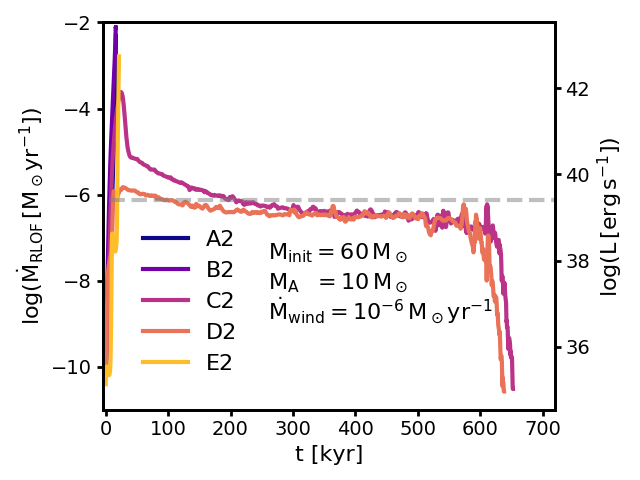}
\includegraphics[width=0.49\textwidth]{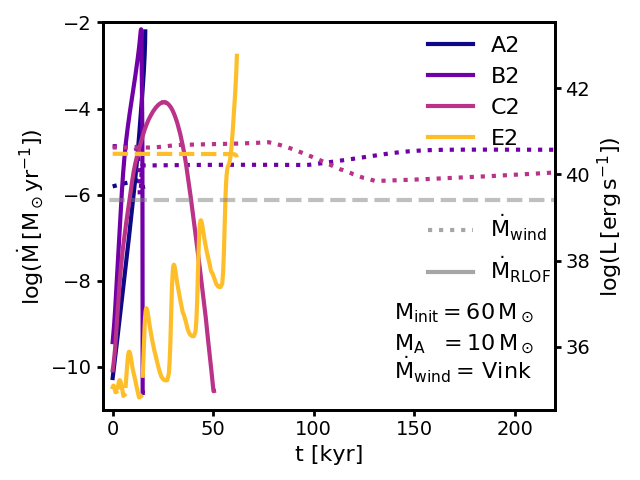}
\caption{Similar to Figure \ref{bin_evo_1_ns} but with a BH accretor, $\eta=0.06$.}
\label{bin_evo_1_bh}
\end{figure*}

In Figs\,\ref{bin_evo_1_ns} and\,\ref{bin_evo_1_bh}, we show the time evolution of the mass-transfer rate
for all our binary evolution models (except for the conservative systems; see Fig.\,\ref{ulx}.

Table\,\ref{tab} gives the key parameters for the of all computed binary models of this paper. 
It also gives the X-ray lifetime $t_\text{X}$, which we  define as the time period during which the accretion luminosity exceeds $10^{33}\,\text{erg/s}$.
To estimate this, we introduce the quantities $L_\text{X,Edd}$ defined as the Eddington limited accretion luminosity 
after half of the X-ray lifetime has passed. We define the quantity $L_\text{X,max}$ in the same manner as $L_\text{X,Edd}$ 
but under the condition that the accretion was not Eddington limited, i.e. as if all of the transferred mass was accreted by the compact object. 
We rather use the luminosity values in the middle of the mass-transfer phase, 
since the arithmetic mean over time would be distorted by the initial peaks of the mass-transfer rate 
which appear in some calculations. In some models these peaks are two orders of magnitude above the long-term mass-transfer rate 
and last for roughly a tenth of the mass-transfer phase. The arithmetic mean would hence be shifted by one order of magnitude 
above the long term X-ray luminosity. Our definition avoids such a distortion and gives a better estimate 
of the typical luminosity during the mass transfer.


\newpage
\begin{landscape}
\begin{table}[htbp!]
~\\
\caption{Summary table of our computed binary evolution models.
All models assume Eddington limited accretion and isotropic re-emission, except for
the last two models, which are computed assuming conservative mass transfer.
The first column indicates the initial donor star model selected for the binary
calculations (cf., Fig.\,\ref{bin_evo_mr}), and
$M_\text{cc,\, init}$ is the initial mass of the compact companion.
$\dot{M}_\text{wind}$ indicates the adopted wind mass loss recipe,
$\frac{\mathrm{d}Y}{\mathrm{d}m}$ measures the helium gradient within the donor star.
$M$,
$L$, and
$T_\text{eff}$ refer to the donor's mass, luminosity and effective temperature, and
$\mathcal{L}=T_\text{eff}^4/g$,
$R$ is the donor's radius,
$Y_\text{S}$ is its surface helium abundance,
$P$ is the binary's orbital period,
$t_\text{X}$ is the X-ray lifetime, and
$L_\text{X,Edd}$ and
$L_\text{X,max}$
are the X-ray luminosities in the middle of the mass-transfer phase assuming Eddington limited
accretion or conservative mass transfer onto the compact object, respectively (see text).
Subscripts {\em init} and {\em final} refer to the times of the onset of mass transfer,
and the end of the calculations, respectively.
}
\resizebox{24cm}{!}{
\begin{tabular}{*{22}{r}} \hline

\rule{0mm}{5mm}Model& $M_\text{CC,\,init}$& $\dot{M}_\text{wind}$& $\frac{\mathrm{d}Y}{\mathrm{d}m}$& $M_\text{init}$&$\log (L_\text{init})$& $T_\text{eff; init}$&$\log (\mathcal{L}_\text{init})$&$R_\text{init}$&$Y_\text{S; init}$&$P_\text{init}$&$t_\text{X}$&$L_\text{X;\,Edd}$&$L_\text{X;\, max}$& $M_\text{final}$&$\log (L_\text{final})$& $T_\text{eff; final}$&$\log (\mathcal{L}_\text{final})$&$R_\text{final}$&$Y_\text{S; final}$&$P_\text{final}$\\ 

\rule{0mm}{5mm}& $[\text{M}_\sun]$& $[\text{M}_\sun \text{yr}^{-1}]$& $[\text{M}^{-1}_\sun]$& $[\text{M}_\sun]$&$[\text{L}_\sun]$&$[\text{kK}]$&$[\mathcal{L}_\sun]$&$[\text{R}_\sun]$&&$[\text{d}]$&$[\text{kyr}]$&$[\text{erg/s}]$&$[\text{erg/s}]$& $[\text{M}_\sun]$&$[\text{L}_\sun]$&$[\text{kK}]$&$[\mathcal{L}_\sun]$&$[\text{R}_\sun]$&&$[\text{d}]$\\ \hline

\rule {0mm}{5mm}
A1&	10&	-&	0.04&	50.0&	5.88&	 24&	4.18&	 49&	0.26&	14.1&	 13&	$2.5\times 10^{39}$&	$1.9\times 10^{40}$&	38.4&	5.29&	 38&	3.70&	 10&	0.48&	1.5\\
B1&	10&	-&	0.04&	42.5&	5.85&	 22&	4.22&	 53&	0.28&	17.9&	484&	$2.5\times 10^{39}$&	$7.5\times 10^{40}$&	30.9&	5.79&	 45&	4.30&	 13&	0.80&	2.5\\
C1&	10&	-&	0.04&	36.5&	5.80&	 41&	4.24&	 16&	0.58&	3.2&	152&	$2.5\times 10^{39}$&	$7.9\times 10^{40}$&	31.7&	5.73&	 52&	4.23&	  9&	0.80&	1.4\\
D1&	10&	-&	0.04&	33.5&	5.75&	 50&	4.23&	 10&	0.72&	1.7&	 29&	$2.5\times 10^{39}$&	$8.0\times 10^{40}$&	31.8&	5.70&	 53&	4.20&	  8&	0.80&	1.3\\
E1&	10&	-&	0.04&	30.5&	5.69&	 53&	4.20&	  8&	0.80&	1.3&	 20&	$7.2\times 10^{38}$&	$7.2\times 10^{38}$&	30.0&	5.65&	 53&	4.17&	  8&	0.80&	1.2\\[2mm]

A1&	2&	-&	0.04&	50.0&	5.88&	 24&	4.18&	 49&	0.26&	11.0&	  7&	$3.1\times 10^{38}$&	$4.2\times 10^{38}$&	49.3&	5.73&	 30&	4.04&	 26&	0.26&	4.0\\
B1&	2&	-&	0.04&	42.5&	5.85&	 22&	4.22&	 53&	0.28&	13.9&	  9&	$3.1\times 10^{38}$&	$1.5\times 10^{39}$&	41.5&	5.75&	 33&	4.13&	 23&	0.32&	3.7\\
C1&	2&	-&	0.04&	36.5&	5.80&	 41&	4.24&	 16&	0.58&	2.5&	  8&	$3.1\times 10^{38}$&	$5.6\times 10^{38}$&	36.3&	5.77&	 43&	4.21&	 14&	0.58&	1.9\\
D1&	2&	-&	0.04&	33.5&	5.75&	 50&	4.23&	 10&	0.72&	1.3&	 10&	$1.4\times 10^{37}$&	$1.4\times 10^{37}$&	33.5&	5.75&	 50&	4.22&	 10&	0.73&	1.2\\
E1&	2&	-&	0.04&	30.5&	5.69&	 53&	4.20&	  8&	0.80&	1.0&	 15&	$4.1\times 10^{36}$&	$4.1\times 10^{36}$&	30.5&	5.68&	 53&	4.20&	  8&	0.80&	1.0\\[2mm]

A2&	10&	-&	0.4&	50.0&	5.83&	 27&	4.13&	 35&	0.26&	8.7&	 13&	$2.5\times 10^{39}$&	$3.0\times 10^{39}$&	45.4&	5.49&	 31&	3.83&	 19&	0.26&	3.4\\
B2&	10&	-&	0.4&	42.5&	5.79&	 22&	4.16&	 52&	0.26&	17.3&	 17&	$2.5\times 10^{39}$&	$2.8\times 10^{40}$&	33.7&	5.61&	 35&	4.08&	 17&	0.27&	3.6\\
C2&	10&	-&	0.4&	36.5&	5.76&	 19&	4.20&	 66&	0.26&	27.5&	674&	$2.6\times 10^{39}$&	$4.9\times 10^{39}$&	32.7&	5.86&	 26&	4.34&	 40&	0.55&	14.1\\
D2&	10&	-&	0.4&	33.5&	5.74&	 26&	4.22&	 36&	0.27&	11.7&	665&	$2.6\times 10^{39}$&	$4.5\times 10^{39}$&	32.5&	5.86&	 30&	4.35&	 31&	0.59&	9.8\\
E2&	10&	-&	0.4&	30.5&	5.69&	 53&	4.20&	  8&	0.80&	1.3&	 20&	$7.1\times 10^{38}$&	$7.1\times 10^{38}$&	30.0&	5.65&	 53&	4.17&	  8&	0.80&	1.2\\[2mm]

A2&	2&	-&	0.4&	50.0&	5.83&	 27&	4.13&	 35&	0.26&	6.8&	  8&	$3.1\times 10^{38}$&	$3.1\times 10^{38}$&	49.8&	5.78&	 29&	4.09&	 30&	0.26&	4.9\\
B2&	2&	-&	0.4&	42.5&	5.79&	 22&	4.16&	 52&	0.26&	13.4&	  9&	$3.1\times 10^{38}$&	$3.1\times 10^{38}$&	41.2&	6.11&	 47&	4.49&	 17&	0.26&	2.3\\
C2&	2&	-&	0.4&	36.5&	5.76&	 19&	4.20&	 66&	0.26&	21.3&	 13&	$2.8\times 10^{38}$&	$2.8\times 10^{38}$&	35.2&	5.59&	 32&	4.04&	 20&	0.26&	3.3\\
D2&	2&	-&	0.4&	33.5&	5.74&	 26&	4.22&	 36&	0.27&	9.1&	264&	$3.1\times 10^{38}$&	$2.5\times 10^{40}$&	32.1&	5.76&	 49&	4.25&	 10&	0.74&	1.4\\
E2&	2&	-&	0.4&	30.5&	5.69&	 53&	4.20&	  8&	0.80&	1.0&	 15&	$2.1\times 10^{36}$&	$2.1\times 10^{36}$&	30.5&	5.68&	 53&	4.19&	  8&	0.80&	1.0\\[2mm]

A2&	10&	$10^{-6}$&	0.4&	50.0&	5.83&	 27&	4.13&	 35&	0.26&	8.7&	 16&	$3.5\times 10^{38}$&	$3.5\times 10^{38}$&	48.0&	5.74&	 30&	4.06&	 27&	0.26&	5.7\\
B2&	10&	$10^{-6}$&	0.4&	42.5&	5.79&	 22&	4.16&	 52&	0.26&	17.3&	 16&	$2.5\times 10^{39}$&	$3.9\times 10^{40}$&	33.4&	5.62&	 36&	4.10&	 17&	0.27&	3.4\\
C2&	10&	$10^{-6}$&	0.4&	36.5&	5.76&	 19&	4.20&	 66&	0.26&	27.5&	652&	$1.5\times 10^{39}$&	$1.5\times 10^{39}$&	32.7&	5.86&	 25&	4.34&	 42&	0.54&	16.4\\
D2&	10&	$10^{-6}$&	0.4&	33.5&	5.74&	 26&	4.22&	 36&	0.27&	11.7&	637&	$1.3\times 10^{39}$&	$1.2\times 10^{39}$&	32.6&	5.86&	 29&	4.34&	 32&	0.59&	11.4\\
E2&	10&	$10^{-6}$&	0.4&	30.5&	5.69&	 53&	4.20&	  8&	0.80&	1.3&	 21&	$4.3\times 10^{38}$&	$4.3\times 10^{38}$&	29.9&	5.64&	 53&	4.17&	  8&	0.80&	1.2\\[2mm]

A2&	10&	Vink&	0.4&	50.0&	5.83&	 27&	4.13&	 35&	0.26&	8.7&	 16&	$3.4\times 10^{38}$&	$3.3\times 10^{38}$&	47.7&	5.73&	 30&	4.05&	 26&	0.26&	5.5\\
B2&	10&	Vink&	0.4&	42.5&	5.79&	 22&	4.16&	 52&	0.26&	17.3&	 14&	$2.5\times 10^{39}$&	$1.2\times 10^{41}$&	25.9&	5.80&	 47&	4.39&	 12&	0.80&	3.6\\
C2&	10&	Vink&	0.4&	36.5&	5.76&	 19&	4.20&	 66&	0.26&	27.5&	 50&	$2.5\times 10^{39}$&	$4.8\times 10^{41}$&	28.2&	5.85&	 45&	4.40&	 14&	0.80&	22.6\\
E2&	10&	Vink&	0.4&	30.5&	5.69&	 53&	4.20&	  8&	0.80&	1.3&	 59&	$1.0\times 10^{38}$&	$4.7\times 10^{37}$&	28.5&	5.62&	 52&	4.16&	  8&	0.80&	1.2\\[2mm]

A2&	2&	$10^{-6}$&	0.4&	50.0&	5.83&	 27&	4.13&	 35&	0.26&	6.8&	 12&	$8.7\times 10^{37}$&	$8.5\times 10^{37}$&	49.9&	5.82&	 28&	4.12&	 34&	0.26&	6.1\\
B2&	2&	$10^{-6}$&	0.4&	42.5&	5.79&	 22&	4.16&	 52&	0.26&	13.4&	  8&	$4.1\times 10^{38}$&	$4.1\times 10^{38}$&	42.0&	5.71&	 26&	4.08&	 33&	0.26&	6.4\\
C2&	2&	$10^{-6}$&	0.4&	36.5&	5.76&	 19&	4.20&	 66&	0.26&	21.3&	 13&	$4.3\times 10^{38}$&	$3.9\times 10^{38}$&	35.3&	5.63&	 30&	4.08&	 23&	0.26&	4.2\\
D2&	2&	$10^{-6}$&	0.4&	33.5&	5.74&	 26&	4.22&	 36&	0.27&	9.1&	639&	$1.3\times 10^{39}$&	$5.3\times 10^{39}$&	32.3&	5.87&	 38&	4.36&	 20&	0.67&	4.9\\
E2&	2&	$10^{-6}$&	0.4&	30.5&	5.69&	 53&	4.20&	  8&	0.80&	1.0&	 18&	$2.1\times 10^{38}$&	$2.1\times 10^{38}$&	30.3&	5.63&	 54&	4.15&	  7&	0.80&	0.9\\[2mm]

A2&	2&	Vink&	0.4&	50.0&	5.83&	 27&	4.13&	 35&	0.26&	6.8&	 11&	$7.4\times 10^{37}$&	$7.1\times 10^{37}$&	49.9&	5.83&	 27&	4.13&	 36&	0.26&	6.7\\
B2&	2&	Vink&	0.4&	42.5&	5.79&	 22&	4.16&	 52&	0.26&	13.4&	  7&	$4.2\times 10^{38}$&	$4.0\times 10^{38}$&	41.7&	5.69&	 27&	4.07&	 30&	0.26&	5.7\\
C2&	2&	Vink&	0.4&	36.5&	5.76&	 19&	4.20&	 66&	0.26&	21.3&	 14&	$1.1\times 10^{39}$&	$6.2\times 10^{38}$&	35.1&	5.64&	 30&	4.10&	 24&	0.26&	4.4\\
E2&	2&	Vink&	0.4&	30.5&	5.69&	 53&	4.20&	  8&	0.80&	1.0&	 85&	$2.8\times 10^{37}$&	$8.3\times 10^{35}$&	28.8&	5.61&	 53&	4.15&	  8&	0.80&	0.9\\[2mm] 

D2 cons.&	10&	-&	0.4&	33.5&	5.74&	 26&	4.22&	 36&	0.27&	11.7&	664&	$4.8\times 10^{39}$&	$4.8\times 10^{39}$&	32.5&	5.86&	 30&	4.34&	 31&	0.59&	9.7\\
D2 cons.&	2&	-&	0.4&	33.5&	5.74&	 26&	4.22&	 36&	0.27&	9.1&	431&	$9.1\times 10^{39}$&	$9.1\times 10^{39}$&	31.8&	5.77&	 48&	4.27&	 11&	0.80&	1.6\\\hline
\end{tabular}}

\label{tab}
\end{table}
\end{landscape}

\end{document}